\documentclass[preprint,12pt,authoryear]{elsarticle}

\usepackage{amssymb}
\usepackage{xcolor}
\usepackage{adjustbox}
\usepackage{caption}
\usepackage{subcaption}
\usepackage{amsmath}
\usepackage{graphbox}
\usepackage{booktabs}
\usepackage{multirow}

\usepackage[hidelinks]{hyperref}
\usepackage{lineno}

\makeatletter
\def\ps@pprintTitle{%
  \let\@oddhead\@empty
  \let\@evenhead\@empty
  \let\@oddfoot\@empty
  \let\@evenfoot\@oddfoot
}
\makeatother

\begin{document}

\begin{frontmatter}

\title{Deep Learning Hyperspectral Pansharpening on large scale PRISMA dataset}

\author[unimib,infn]{Simone Zini\corref{cor}}
\ead{simone.zini@unimib.it}
\author[unimib]{Mirko Paolo Barbato}
\ead{mirko.barbato@unimib.it}
\author[unimib]{Flavio Piccoli}
\ead{flavio.piccoli@unimib.it}
\author[unimib]{Paolo Napoletano}
\ead{paolo.napoletano@unimib.it}

\cortext[cor]{Corresponding author}
\affiliation[unimib]{organization={Imaging and Vision Laboratory, Department of Informatics, Systems and Communication, University of Milano-Bicocca},
            addressline={Viale Sarca 336},
            city={Milano},
            postcode={20126},
            country={Italy}}

\affiliation[infn]{organization={Istituto Nazionale di Fisica Nucleare (INFN)},
            city={Milano},
            postcode={20126},
            country={Italy}}

\begin{abstract}
In this work, we assess several deep learning strategies for hyperspectral pansharpening. 
First, we present a new dataset with a greater extent than any other in the state of the art.
This dataset, collected using the ASI PRISMA satellite, covers about 262200 $km^2$, and its heterogeneity is granted by randomly sampling the Earth's soil. Second, 
we adapted several state of the art approaches based on deep learning to fit PRISMA hyperspectral data and then assessed, quantitatively and qualitatively, the performance in this new scenario.
The investigation has included two settings: Reduced Resolution (RR) to evaluate the techniques in a supervised environment and Full Resolution (FR) for a real-world evaluation. 
The main purpose is the evaluation of the reconstruction fidelity of the considered methods.
In both scenarios, for the sake of completeness, we also included machine-learning-free approaches. From this extensive analysis has emerged that data-driven neural network methods outperform machine-learning-free approaches and adapt better to the task of hyperspectral pansharpening, both in RR and FR protocols. 

\end{abstract}


\end{frontmatter}

\section{Introduction}

Remote Sensing (RS) has revolutionized our ability to observe and analyze our planet from a vantage point beyond the Earth's surface~\citep{salcedo2020machine}. The analysis of data gathered by sensors on board of satellites or aircraft, in fact, allows the inference of useful information about the land, water and atmospheric systems of the Earth. This technology became fundamental in several fields, such as environmental monitoring~\citep{barbato2022unsupervised}, agriculture~\citep{sishodia2020applications}, urban planning~\citep{wellmann2020remote}, disaster management \citep{van2000remote}, and resource exploration~\citep{frick2019framework}.

However, the costs to put a satellite in Earth orbit are very high. They range from 2.6k\$/kg with SpaceX to 22k\$/kg with NASA, with an intermediate value of 17.6k\$/kg with Soyuz, the Russian rockets~\citep{costs2002trends,jones2018recent}. Minimizing the payload is therefore the major goal that drives the choice and the design of every component on a satellite \citep{okninski2021hybrid}.
This constraint, in combination with the need to use as little energy as possible, results in a huge tradeoff between the spatial resolution and the number of acquired bands when designing optical remote sensing devices. On the one hand, in fact, several orbital expeditions such as Landsat 6/7 \citep{wulder2019current}, SPOT 6/7 \citep{chevrel1981spot}, Sentinel-2 \citep{phiri2020sentinel} included a panchromatic imaging device acquiring at high resolution~\citep{apostolopoulos2022spot}. On the other hand, missions carrying hyperspectral imaging devices such as ASI PRISMA~\footnote{\url{https://www.asi.it/scienze-della-terra/prisma/}}, had to decrease the spatial resolution in favor of a higher number of acquired bands \citep{krueger2010closesat}.

The loss of spatial resolution can be partially solved through the use of pansharpening \citep{li2017pixel}. In this context, the panchromatic image could be used as a source of information to extend the spatial resolution of the multispectral (MS) and hyperspectral (HS) images.

The first attempts of image pansharpening are machine-learning-free approaches \citep{vivone2014critical}, designed to handle data in the range of visible radiations (400 - 700 nm). These approaches assume the possibility to exploit the existing relation between the panchromatic image and the spectral bands in the input data.
This assumption may not be valid when handling data outside the range of visible wavelengths, i.e. when there is partial or missing spectral overlap among the panchromatic image and the spectral bands to be processed.

Alongside these methods, neural network-based approaches have been recently developed, showing promising results.
However, data-driven approaches are limited by the lack of high-cardinality datasets in the state of the art. 
Table~\ref{tab:datasets} reports the most relevant datasets in the literature used for multispectral and hyperspectral RS pansharpening. The majority of them are composed of only one single satellite or aerial image covering a small portion of land (at most few $km^2$), with limited variability in the content of the scene.

In order to overcome the limitations relative to the lack of data, in this paper we present:
\begin{itemize}
    \item a new large-scale dataset covering 262200 $km^2$ for qualitative assessment of deep neural models for hyperspectral image pansharpening. Such dataset is collected from the PRISMA satellite, preprocessed and adopted for the retraining of current state-of-the-art approaches for image pansharpening;

    \item an in-depth comparison, both in quantitative and qualitative terms, of the current deep learning approaches, re-trained and tested on the newly proposed large scale dataset, and traditional machine-learning-free approaches.
    
\end{itemize}

To the best of our knowledge, the presented study is the first one based on a large scale dataset, covering a big variety of ground areas. We believe that the proposed investigation will be a starting point for the design of new deep learning approaches for RS image pansharpening.

\section{Related works} 

The field of pansharpening has made significant advancements thanks to data-driven approaches and novel methods. This section focuses on the current state of the art in pansharpening, highlighting two important aspects: the development of benchmarks for a comprehensive evaluation and the design of novel methods that effectively improve spatial resolution while maintaining spectral fidelity. By reviewing these benchmarks and methods, we aim to provide a comprehensive overview of the most recent advancements. 

\subsection{Datasets}
\label{sub:benchmarks}

\begin{table}[t]
    \centering
    \caption{List of existing datasets used for RS image pansharpening. Here are reported, for each dataset, the number of images of the dataset, the number of bands, and the coverage in terms of wavelength. {The image resolutions reported in this table are taken from the original dataset descriptions.}}

    \adjustbox{width=\linewidth}{
    \begin{tabular}{lrrrrr}
    \hline
         Dataset & \multicolumn{1}{l}{Cardinality} & \multicolumn{1}{l}{Images resolution} & \multicolumn{1}{l}{Type}& \multicolumn{1}{l}{\# of bands} & \multicolumn{1}{l}{Wavelength coverage}  \\\hline
         
         Pavia University       & 1 & 610 $\times$ 610 & airborne & 103 & 430 - 838 nm \\
         Pavia Center           & 1 & 1096 $\times$ 1096& airborne & 102 & 430 - 860 nm \\
         Houston \citep{he2023dynamic, labate2019structured} & 1 & 349 $\times$ 1905 & airborne & 144 & 364 - 1046 nm \\
         Chikusei \citep{yokoya2016airborne}               & 1 & 2517 $\times$ 2335 & airborne & 128 & 363 - 1018 nm \\
         AVIRIS Moffett Field \citep{loncan2015hyperspectral} & 1 & 37 $\times$ 79 & airborne & 224 & 400 - 2500 nm \\
         Garons \citep{loncan2015hyperspectral} & 1 & 80 $\times$ 80 & airborne & 125 & 400 - 2500 nm \\
         Camargue \citep{loncan2015hyperspectral} & 1 & 100 $\times$ 100 & airborne & 125 & 400 - 2500 nm \\
         Indian Pines~\citep{baumgardner2015220}           & 1 & 145 $\times$ 145 & airborne & 224 & 400 - 2500 nm \\
         Cuprite Mine           & 1 & 400 $\times$ 350 & airborne & 185 & 400 - 2450 nm \\
         Salinas & 1 & 512 $\times$ 217 & airborne & 202 (224) & 400 - 2500 nm \\
         Washington Mall \citep{he2019hyperpnn} & 1 & 1200 $\times$ 300 & airborne & 191 (210) & 400 - 2400 nm \\
         Merced \citep{he2023dynamic} & 1 & 180 $\times$ 180 & satellite & 134 (242) & 400 - 2500 nm \\
         Halls Creek \citep{he2020spectral} & 1 & 3483 $\times$ 567 & satellite & 171 (230) & 400 - 2500 nm \\
        \textbf{OURS based on PRISMA}   & 190 & 1259 $\times$ 1225 & satellite & 203 (230)& 400 - 2505 nm \\\hline
    \end{tabular}
    }
    \label{tab:datasets}
\end{table}

The choice and characteristics of the benchmark datasets play a crucial role in evaluating and comparing different algorithms for RS pansharpening. Each dataset has unique properties that differentiate it from others, including cardinality, image resolution, acquisition setup, number of spectral bands, and wavelength coverage. 
For an immediate comparison of the existing datasets in the state of the art, please refer to Table \ref{tab:datasets} which summarizes all of them and their properties.

It is possible to divide the existing datasets into different groups by considering mainly three properties: the wavelength coverage, image resolution, and acquisition setup of each dataset.
Regarding the wavelength coverage, four datasets are in the range from visible to near infra-red (VNIR), while the remaining ones cover the entire spectrum, from visible to short-wave infrared (SWIR).
The use of data with limited spectral coverage for the design of pansharpening algorithms could limit their applicability to real case scenarios, which may require the use of bands and data not covered by those datasets.

Another important aspect is the image resolution, associated to the dataset cardinality.
While datasets like Halls Creek \citep{he2020spectral} can potentially be tiled in smaller samples for training or validation purposes, the other ones are limited due to the low cardinality and low resolution of the data. Futhermore, even if an image is tiled, the variety of the content of the scenes considered is limited to the area covered by the single image, making it hard to evaluate algorithms in different scenarios. Finally, most of the datasets in the state of the art are tagged as ``airborne" type, which means that are collected by using airplanes or low-altitude flying devices, while only two are made of satellite-collected images.

\subsection{Methods}
\label{sub:methods}

\newcommand{\sotaspace}[0]{\vspace{3mm}}
 
Pansharpening methods can be grouped into five categories: component substitution (CS), multiresolution analysis (MRA), Bayesian, matrix factorization (as defined by \cite{loncan2015hyperspectral}) and deep learning.
\newline

\noindent\textbf{Component substitution methods} consist of substituting the spatial component of the spectral images with the high-resolution panchromatic images. The results are obtained by projecting a high-resolution version of the spectral image into its spatial component and then reverting the transformation using the panchromatic information instead of the spatial projection extracted. CS includes methods such as principal component
analysis (PCA) \citep{chavez1991comparison}, intensity-hue-saturation (IHS) \citep{tu2004fast}, Gram-Schmidt (GS) \citep{laben2000process}, and GS Adaptive (GSA) \citep{4305344}. These methods are usually easy to implement, achieve high spatial fidelity, and are robust to misregistration, but can create significant spectral distortions \citep{loncan2015hyperspectral}.

\sotaspace

\noindent\textbf{In multiresolution analysis}, \cite{loncan2015hyperspectral} includes methods such as Decimated Wavelet Transform (DWT) \citep{mallat1989theory}, Undecimated Wavelet Transform (UDWT) \citep{nason1995stationary}, “à-trous” wavelet transform (ATWT) \citep{shensa1992discrete}, and Laplacian pyramid~\citep{burt1987laplacian}, which consist of using a filtered version of the PAN signal, to extract high-resolution details and inject them into the spectral image. Compared to the CS methods, the MRA techniques are more difficult to implement and computationally more complex, but also allow for achieving a better spectral consistency with the original spectral information.

Some other approaches, for instance, Guided Filter PCA~\citep{liao2015processing}, combine the two techniques to gather the advantages from both, but the results on hyperspectral images were not promising, being the technique with the worst results on hyperspectral data in \cite{loncan2015hyperspectral}'s investigation.

\sotaspace
\noindent\textbf{Bayesian approaches} are based on the estimation of the posterior probability of the full-resolution image that would be obtained considering the original panchromatic and spectral information. These methods typically consider the sensor characteristics to enhance the resolution, thus achieving good results but also being less generalizable and more complex to use \citep{loncan2015hyperspectral}.

\sotaspace
\noindent\textbf{Matrix factorization techniques} are described by \cite{loncan2015hyperspectral} as the only ones purposely used for hyperspectral pansharpening, and instead of using the panchromatic information, use a high-resolution multispectral image to convey the spatial information of the hyperspectral data into a higher resolution space. Even in this case, to exploit the best factorization to reconstruct the new image, the characteristics of the sensors are taken into consideration, making them less viable compared to other methods.

\sotaspace
\noindent\textbf{Deep Learning approaches}, due to the recent success of the Neural Networks, have been proposed in recent years and using different models for multispectral pansharpening. \citep{zhang2023panchromatic}. 
\cite{masi2016pansharpening} in 2016 proposes the first Convolutional Neural Network designed for multispectral pansharpening. They presented a simple neural network made of few convolutional layers, capable to outperform the machine-learning-free approaches on three standard datasets. {In 2018, \cite{scarpa2018target} tried to compensate for the lack of data, proposing the Adaptive-PNN strategy by fine-tuning the model extracting samples from the reference image, and refining the pansharpened reconstruction by closing the gap between training and testing in pansharpening.}
\cite{yang2017pannet}  proposed a new model with the focus on spectral information preservation, which operates mainly on the high-frequency components of the multispectral images, while trying to keep low frequency information as much as possible unaltered. 
In 2018, \cite{yuan2018multiscale} proposed a multi-branch network, being the first approach to fuse PAN and MS images in feature domain and reconstruct the pan-sharpened image from the fused features, instead of approaching the problem as a super-resolution task.
In 2020, \cite{liu2020remote} presented a multi-resolution based approach for image fusion, based on wavelets decomposition, which represents the first approach to explicitly perform a deep fusion operation between the panchromatic information and the multispectral bands. 
In the same year, \cite{cai2020super} propose a method for progressive pansharpening, while recently, in 2021, \cite{xie2021detail} proposed a progressive PAN-injected fusion method based on super-resolution. This last approach extracts information from the panchromatic image with dedicated encoders branches, from both low and high frequencies, in order to better exploit features from the panchromatic image, achieving state of the art results.

{Some attempts at deep learning approaches for hyperspectral data have also been investigated. In 2019, \cite{he2019hyperpnn} proposed HyperPNN, a phases CNN that firstly extracts spatial features from the panchromatic image and spectral features from the hyperspectral one, secondly fuses the spatial and spectral features with dedicated convolutional layers, and thirdly predicts the spectral information of the pansharpened image with convolutional layers that focus only on the spectral signatures. This model has been followed in 2020 by the improved version  called HySpecNet \citep{he2020spectral}. In the same year of HyperPNN, \cite{zheng2019deep} investigated the use of the residual network for pansharpening, firstly guiding the upscaling and enhancing the edge details of the hyperspectral data with Contrast Limited Adaptive Histogram Equalization (CLAHE) and a guided filter to fuse the image with the panchromatic information, and then using a Deep Residual Convolutional Neural Network (DRCNN) to boost the reconstruction. \cite{xie2019hyperspectral} developed the HS Pansharpening method with Deep Priors (HPDP), exploiting the power of different deep learning modules to improve all parts of the pansharpening pipeline. In particular, they used a Super Resolution Deep Learning (SRDL) module to upscale the HS image and fuse it with the panchromatic information by also considering high-frequency information extracted by the proposed High-Frequency Net (HFNet). They finally obtained the final high-resolution HS by injecting the high-frequency structure in the upscaled HS, using a Sylvester equation. It is worth noticing that they used multispectral images for training to compensate for the limited number of training samples. Recently, in 2023, \cite{he2023dynamic} proposed dynamic hyperspectral pansharpening that uses a learn-to-learn strategy to adapt the pansharpening to the spatial variations of an image.

Despite the increased adoption of CNNs and deep learning in the field of pansharpening, and the increased interest in the use of hyperspectral images for satellite image analysis, the limited number of hyperspetral samples is still an issue.
In order to study the impact and the possible advantages of the application of deep neural models on hyperspectral pansharpening, in this work we adapted and retrained these models to hyperspectral data, collected from the PRISMA dataset, and compared with a subset of machine-learning-free approaches.}

\section{Materials and Methods}
\subsection{Data}
\label{sec:data}
\begin{table}[]
    \centering
    \caption{Ranges of wavelengths covered by the panchromatic image and by the hyperspectral cubes VNIR and SWIR, and the corresponding number of bands. The PAN image covers most of the range of the VNIR cube, while the SWIR cube is completely outside of that range.}
    \label{tab:prisma_details}
    \adjustbox{width=\linewidth}{
    \begin{tabular}{ccccc}
        \hline
        \multirow{2}{*}{Cube} & Wavelengths covered & \multirow{2}{*}{\# of bands} & \multicolumn{2}{c}{Resolution} \\
                              & nm                                     &                              & m/px       & pixels            \\\hline
        panchromatic    & 400 – 700                          & 1                            & 5          & 7554 × 7350       \\
        VNIR                  & 400 – 1010                         & 66                           & 30         & 1259 × 1225       \\
        SWIR                  & 920 – 2505                         & 174                          & 30         & 1259 × 1225      \\\hline
   \end{tabular}

    }

\end{table}

\begin{figure}[t]
    \centering
    \includegraphics[width=\textwidth]{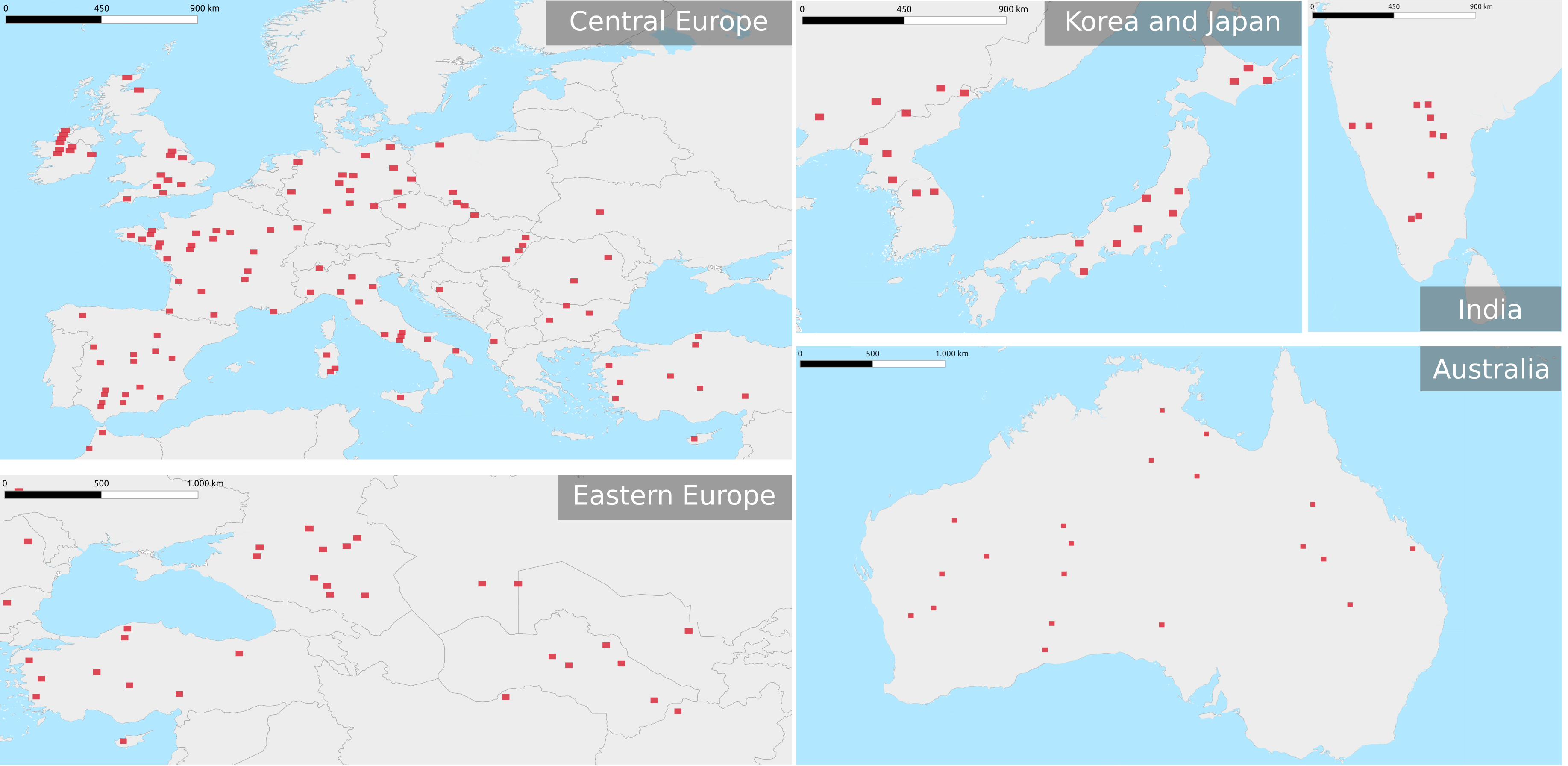}
    \caption{Map of the patches acquired using the PRISMA satellite. On average, every patch covers about $1380\;km^2$ of soil.} 
    \label{fig:worldview}
\end{figure}

\begin{figure}[t]
    \centering
    \begin{tabular}{cccc}
        \multicolumn{4}{c}{$PAN$} \\        
        \includegraphics[width=0.23\textwidth]{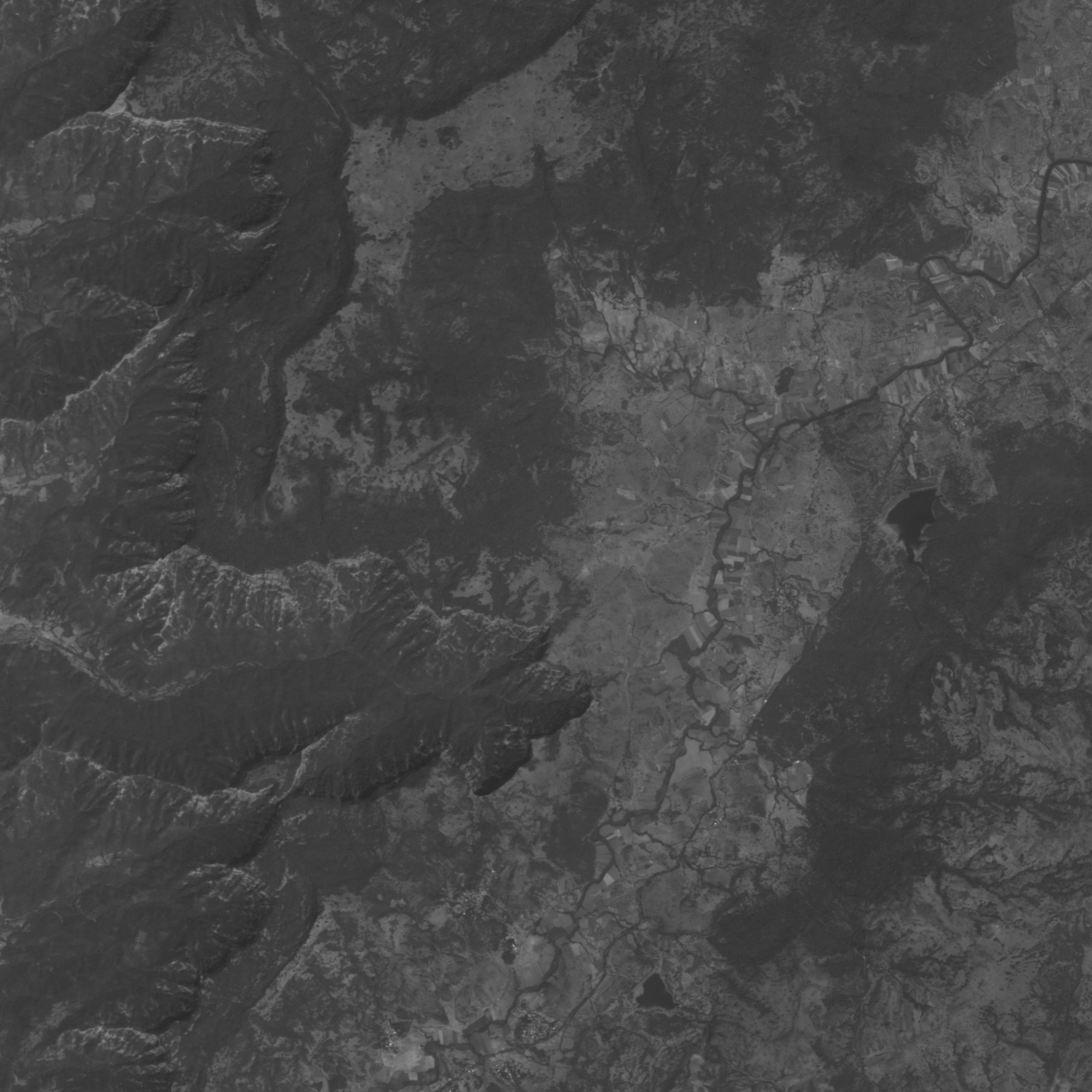} & 
        \includegraphics[width=0.23\textwidth]{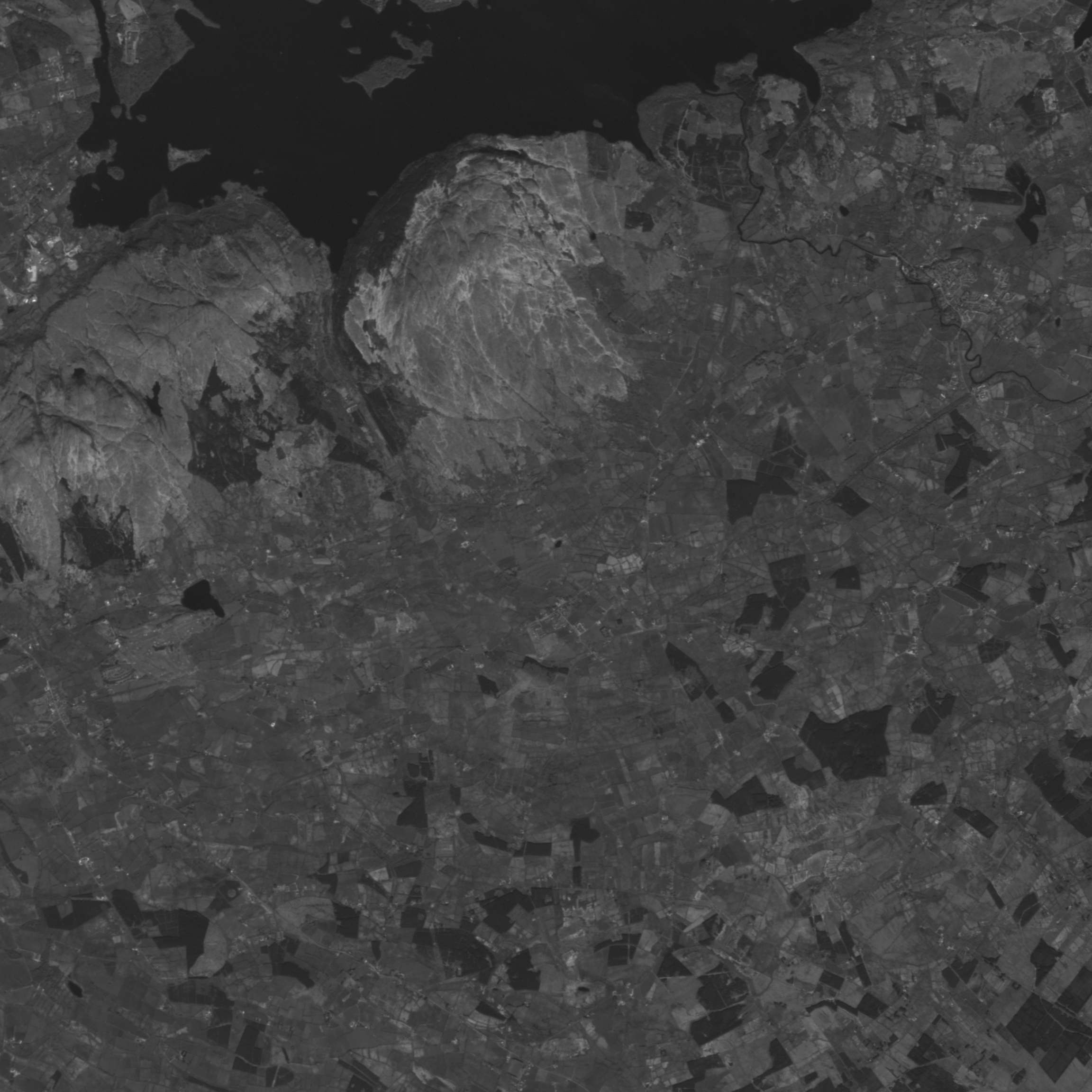} &
        \includegraphics[width=0.23\textwidth]{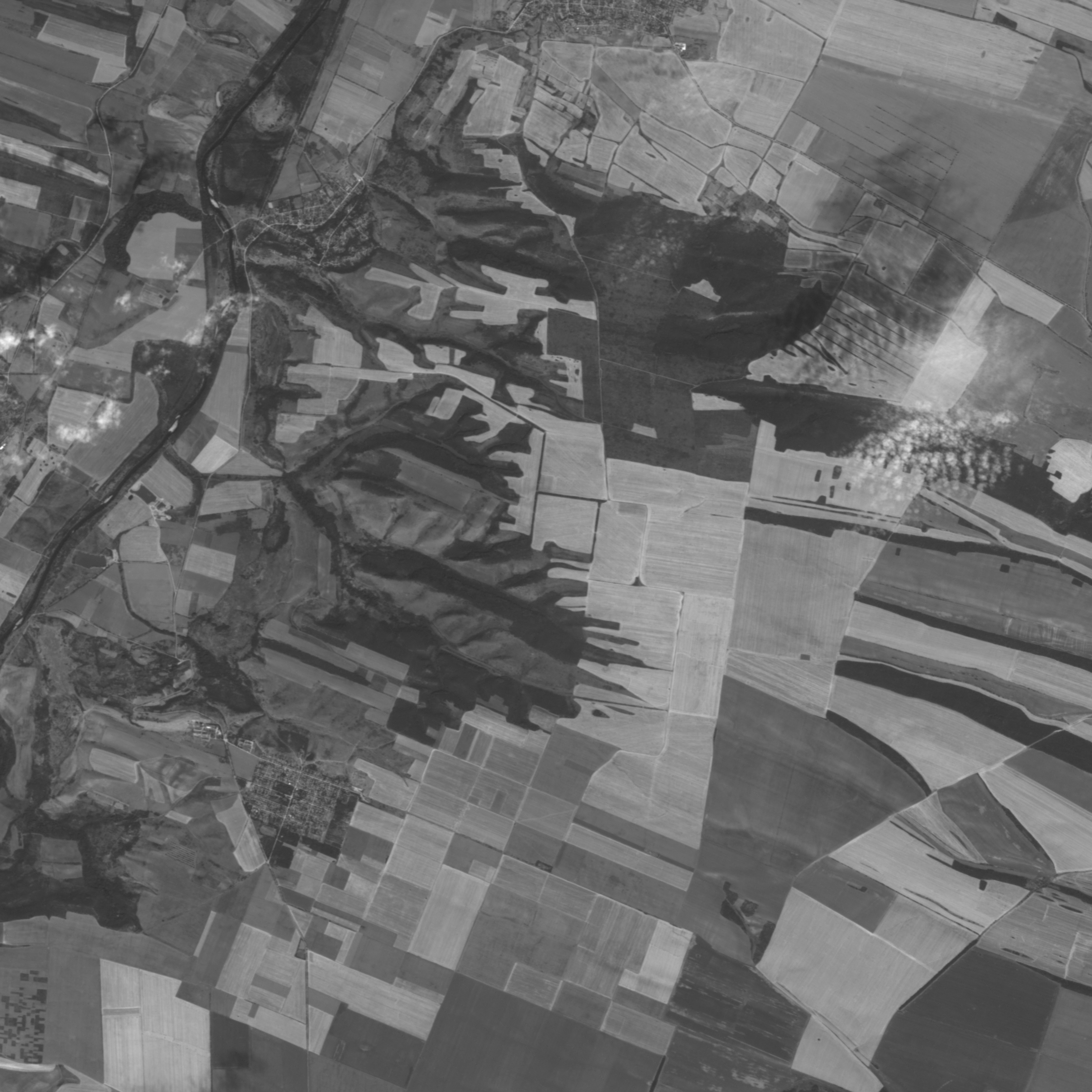} &
        \includegraphics[width=0.23\textwidth]{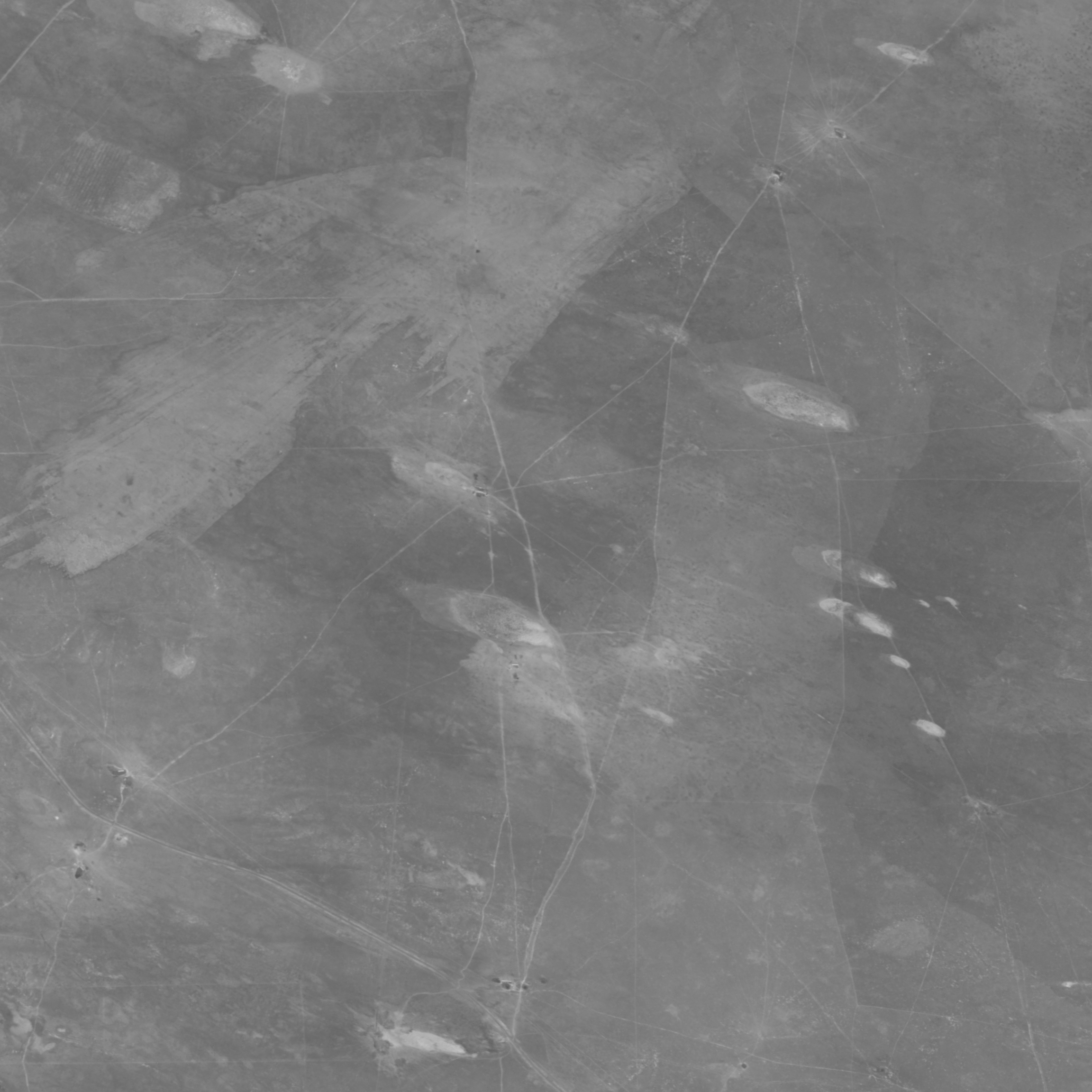} \\

        \multicolumn{4}{c}{$HS$} \\
        \includegraphics[width=0.23\textwidth]{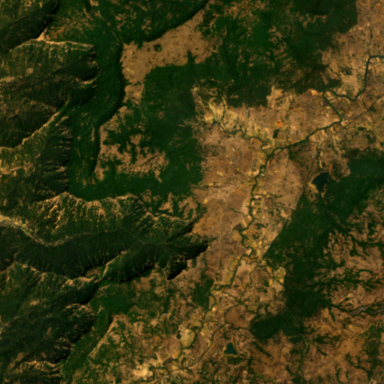} & 
        \includegraphics[width=0.23\textwidth]{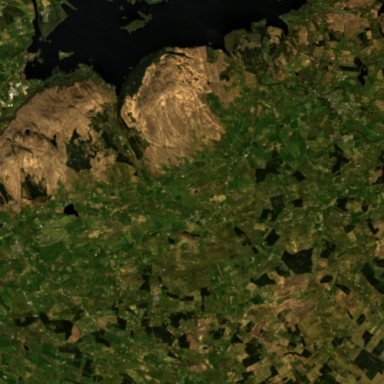} &
        \includegraphics[width=0.23\textwidth]{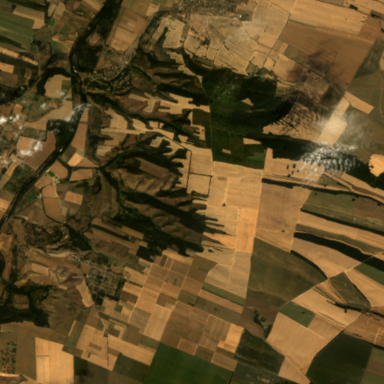} &
        \includegraphics[width=0.23\textwidth]{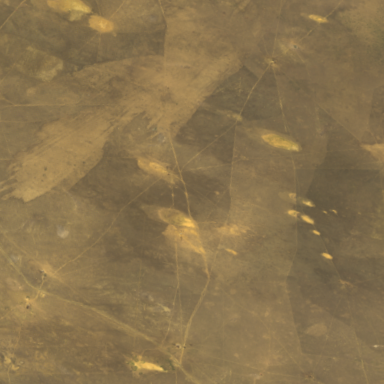} \\

        \multicolumn{4}{c}{$HS_{\downarrow}$} \\
        \includegraphics[width=0.23\textwidth]{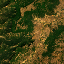} & 
        \includegraphics[width=0.23\textwidth]{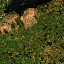} &
        \includegraphics[width=0.23\textwidth]{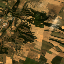} &
        \includegraphics[width=0.23\textwidth]{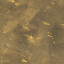} \\

        \end{tabular}
    
    \caption{Examples of PRISMA dataset entries, visualized  in true color RGB ($641~nm$, $563~nm$, $478~nm$). $PAN$ image is at a resolution of 5 meters per pixel, $HS$ images at 30 meters per pixel and $HS_{\downarrow}$ at 180 meters per pixel}
    \label{fig:visual_examples}
\end{figure}

In order to assess the performance of the approaches for hyperspectral image pansharpening, we created a new dataset of HS images, collected using the PRISMA satellite.
We specifically collected $190$ images covering different areas, from Europe, Japan, Korea, India and Australia for a total of about 262200 $km^{2}$. The actual locations of the images are shown in Figure~\ref{fig:worldview}, while in the last row of Table~\ref{tab:datasets} is reported our dataset with a summary of its characteristics, compared with other existing datasets. 

The data used for the construction of the proposed datasets has been collected by using the level-2D image data product downloaded from the ASI PRISMA portal for data distribution\footnote{ASI portal: \url{https://www.asi.it/scienze-della-terra/prisma/}
}.
Visible and Near-InfraRed (VNIR), Short-Wave InfraRed (SWIR) cubes and the panchromatic (PAN) band have been extracted from these downloaded products, according to the Hierarchical Data Format (HDF5) standard.
The hyperspectral cubes from the level-2D refers to the geocoded at-surface (Bottom-of-Atmosphere) reflectance data~\citep{PRISMAdoc}.
PAN images are at a spatial resolution of $5$ meters per pixel, while VNIR and SWIR cubes (respectively 66 and 174 spectral bands) are at a spatial resolution of $30$ meters per pixel. Table~\ref{tab:prisma_details} reports the details of each cube, while Figure~\ref{fig:visual_examples} shows some examples of PRISMA data visualized in true-color RGB. Each PAN image is at a resolution of 7554 $\times$ 7350 pixels, while HS bands are at a resolution of 1259 $\times$ 1225 pixels.

Each collected image has been pre-processed by performing an image co-registration step and a cleaning step, with the last one used 
to remove bands that contain noisy or invalid data. Each image is then divided into tiles at different resolutions, to produce two sets of images for two different training and evaluation protocols: the Full Resolution (FR) protocol and Reduced Resolution (RR) protocol.

\subsubsection{Data cleaning procedure}

\begin{figure}[tb]
    \centering

    \begin{subfigure}[b]{\textwidth}
        \includegraphics[width=.9\linewidth]{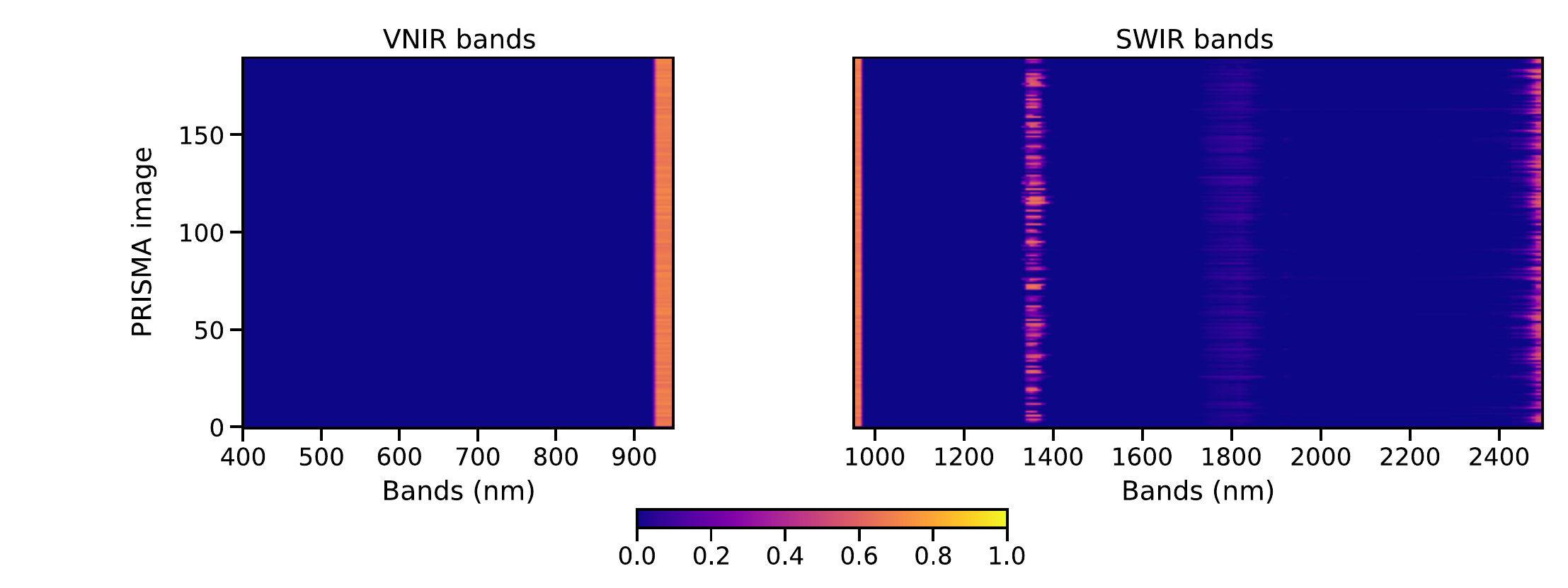}
        \caption{}
        \label{fig:banalysis_a}
    \end{subfigure}

    \begin{subfigure}[b]{\textwidth}
        \includegraphics[width=.9\linewidth]{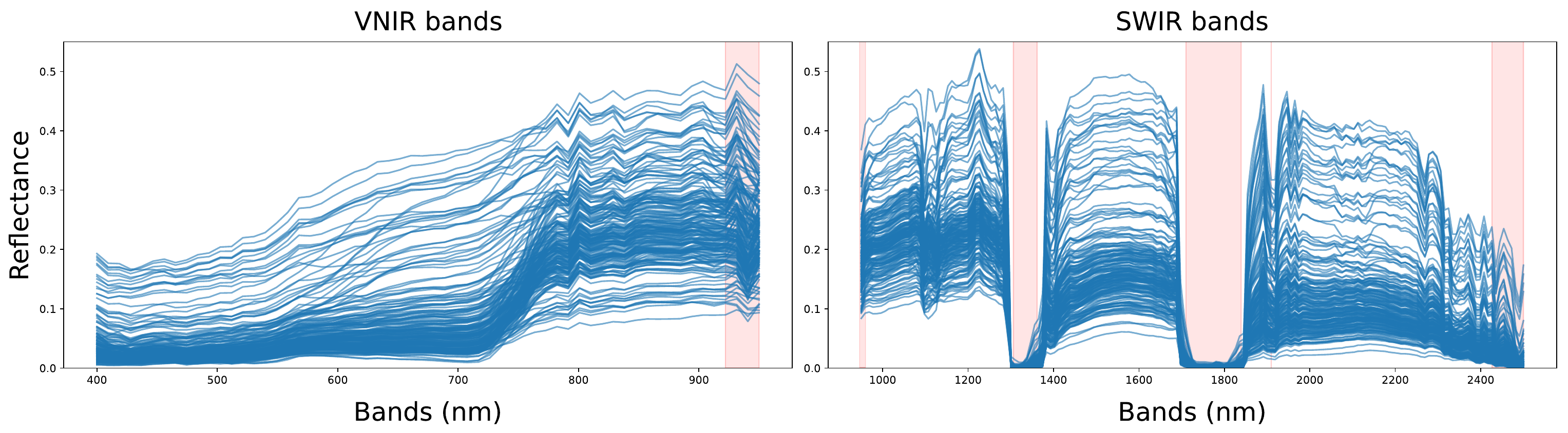}
        \caption{}
        \label{fig:banalysis_b}
    \end{subfigure}
    
    \caption{Distribution of the invalid band for each PRISMA image collected. In (a), the scale indicates the percentage of invalid pixels for each band of each image collected from the PRISMA satellite. Bands that are considered invalid for at least one image (with more than 5\% of the entries invalid), have been selected for removal. In (b) are shown the average spectral signatures per image and the bands excluded in the final version of the dataset.
    }
    \label{fig:banalysis}
\end{figure}

The VNIR and SWIR PRISMA level-2D products cubes cannot be directly used because of two problems:
\begin{itemize}
    \item slight misalignment between panchromatic image and VNIR and SWIR cubes (VNIR and SWIR are assumed to be aligned already);
    \item presence of pixels marked as invalid from the Level-2D Prisma pre-processing.
\end{itemize}

To tackle the first problem,
we adopted the AROSIC framework \citep{scheffler2017arosics} to align the VNIR and SWIR cubes to the corresponding panchromatic images. We manually selected a reference band for the VNIR and SWIR cube to be used for the calculation of the transformation.
The same alignment has been used for all the $240$ bands of VNIR and SWIR cube.

Invalid bands are removed through a cleaning procedure. 
From the PRISMA HDF5 data, we also extracted the VNIR\_PIXEL\_L2\_ERROR and SWIR\_PIXEL\_L2\_ERROR matrices.
These matrices contain pixel-specific annotations regarding the status of the information collected by the satellite.
We removed bands having at least $5\%$ of the pixels labeled as \textsc{invalid}.
More details on the labeling system are available at \cite{PRISMAdoc}.
The selected bands have been removed from all the scenes collected from PRISMA. Figure~\ref{fig:banalysis_a} 
shows the distribution of the invalid bands (x-axis) over all the $190$ selected PRISMA images (y-axis). 
Figure~\ref{fig:banalysis_b}
shows the average spectral signature for each image (blue lines) and the spectral bands that have been removed (pink stripes).
After this cleaning procedure, VNIR and SWIR bands are concatenated, thus obtaining a final HS cube of 203 spectral bands.

\subsubsection{Full Resolution and Reduced Resolution datasets}
\label{sec:dataFRRR}
Our experimentation is made adopting two different protocols which require two different versions of the dataset:
\begin{enumerate}
    \item \textit{Full Resolution (FR)}:  this dataset is used to evaluate the goodness of pansharpening algorithms without a reference image.
    Due to missing reference images, this dataset cannot be used for model training but only for evaluation purpose. This dataset is made of couples of the type $<PAN, HS>$.
    \item \textit{Reduced Resolution (RR)}: this dataset is created in order to perform full-reference evaluation since it presents reference bands alongside the input HS and the PAN, and for training the deep learning model. This dataset is made of triplets of the type $<PAN_{\downarrow}, HS_{\downarrow}, HS>$.
    
\end{enumerate}

To create the two versions of the dataset, we tiled and resized the original collected images with different parameters. Table~\ref{tab:data_versions} provides a summary of the characteristics of the two versions.
\begin{table}[]
    \centering
    \caption{Size and resolution of the input PAN images, HS bands, and pansharpened outputs in both RR and FR protocols.}
    \label{tab:data_versions}
    \adjustbox{width=.7\linewidth}{
    \begin{tabular}{ccccc}
    \toprule
                       & \multirow{2}{*}{Size (px)} & \multirow{2}{*}{Resolution (m/px)} & \multicolumn{2}{c}{Usage}    \\
                       &                                 &                                    & FR & RR        \\\midrule
    $PAN$              & $2304 \times 2304$              & 5                                  & input                  & -         \\
    $PAN_{\downarrow}$ & $384\times384$                  & 30                                 & -                      & input     \\
    $HS$               & $384\times384$                  & 30                                 & input                  & reference \\
    $HS_{\downarrow}$  & $64\times64$                    & 180                                & -                      & input     \\
    \midrule
    $\hat{HS}_{FR}$  & $2304\times2304$                    & 5                                & output                      & -     \\
    $\hat{HS}_{RR}$  & $348\times348$                    & 30                                & -                      & output     \\
    \bottomrule
    \end{tabular}
    }
\end{table}
The FR dataset is made of tiles of size $2304\times2304$ at the original spatial resolution of $5m/px$, for the PAN image, and $384\times384$ pixels at $30m/px$ resolution for the HS bands. In our experimental setup, the pansharpening algorithms are used to scale up the HS bands from 30$m/px$ by a factor of $6\times$,
thus obtaining a no-reference reconstruction $\hat{HS}_{FR}$ of HS bands at a size of $2304\times2304$ at a spatial resolution of $5m/px$. 

The RR dataset is obtained by subsampling the FR version and generating triples of the type $<PAN_{\downarrow}, HS_{\downarrow}, HS>$.
Firstly, the VNIR-SWIR bands are tiled at a dimension of $384\times384$ pixels, which corresponds at a resolution of $30~m/px$ ($HS$). These images are used as reference for the evaluation of the algorithms performance.
Then, the same cubes are further reduced at 1/6 of their original resolution, obtaining new tiles at size $64\times64$ at a spatial resolution of $180~m/px$ ($HS_{\downarrow}$) which are the input of the pansharpening algorithm. 
The panchromatic images are also reduced to 1/6 of the original resolution and tiled at size $384\times384$ pixels at a spatial resolution of $30~m/px$ ($PAN_{\downarrow}$), in order to be used as input for the pansharpening operation. 
The pansharpening algorithm is defined as a function that takes as input the pair $<PAN_{\downarrow}, HS_{\downarrow}>$, and it outputs an approximation $\hat{HS}_{RR}$ of the original HS, which is a $6\times$ version of the $HS_{\downarrow}$.

\subsection{Reduced Resolution Metrics}

We used the following evaluation metrics to compare the pansharpened $\hat{HS}_{RR}$ image and the reference $HS$:

\begin{itemize}
    \item ERGAS~\citep{wald2002data} is an error index that tries to propose a global evaluation of the quality of the fused images.
    This metric is based on the $RMSE$ distance between the bands that constitute the fused and the reference images and is computed as:
        \begin{equation}
            RMSE(x,y)=\sqrt{ \frac{1}{m} \sum^m_{j=1} (x_j-y_j)^2 }
        \end{equation}
        
        \begin{equation}
            ERGAS(x,y) = 100\frac{h}{l}\sqrt{\frac{1}{N} \sum_{i=1}^{N}\left(\frac{RMSE(x_i,y_i)}{\mu(y_i)}\right) ^2}
        \end{equation}

        where $x$ and $y$ are the output pansharpened image and  the reference, respectively, $m$ is the number of the pixels in each band, $h$ and $l$ are the spatial resolution of the PAN image and HS image, respectively, $\mu(y_i)$ is the mean of the $i-th$ band of the reference and $N$ is the number of total bands.
        
    \item The Spectral Angle Mapper (SAM)~\citep{yuhas1992discrimination} denotes the absolute value of the angle between two vectors $v$ and $\hat v$.
        \begin{equation}
            SAM(v,\hat{v})=cos^{-1} \frac{<v,\hat{v}>}{||v||_2\cdot||\hat{v}||_2}
        \end{equation}
    where $v$ and $\hat{v}$ are respectively the flattened versions of $\hat{HS}_{RR}$ and $HS$.
    A SAM value of zero denotes complete absence of spectral distortion but possible radiometric distortion (the two vectors are parallel but have different lengths).

    \item The Spatial Correlation Coefficient (SCC)\citep{zhou1998wavelet} is a spatial evaluation index that analyses the difference in high-frequency details between two images. 
        SCC is computed as follows:

        \begin{equation}
            SCC(x,y) = \frac{ \sum^w_{i=1}\sum^h_{j=1}(F(x)_{i,j} - \mu_{F(x)})(F(y)_{i,j} - \mu_{F(y)}) }
            { \sqrt{ \sum_{i=1}^w\sum^h_{j=1}(F(x)_{i,j} - \mu_{F(x)})^2 \sum^w_{i=1}\sum^h_{j=1}(F(y)_{i,j} - \mu_{F(y)})^2 } }
        \end{equation}
        where $\mu_{F(x)}$ and $\mu_{F(y)}$ are the means of $F(x)$ and $F(y)$ respectively and and $w$ and $h$ are the weight and height of an image. $F$ is a filter for the extraction of high-frequency details, defined as follows:
        
        \begin{equation}
            F = 
            \begin{bmatrix}
            -1 & -1 & -1\\
            -1 & 8 & -1 \\
            -1 & -1 & -1\\
            \end{bmatrix}
        \end{equation}
        
    \item The $Q2^n$ index is a generalization of the Universal Quality Index ($UQI$) defined by \cite{wang2002universal} for an image $x$ and a reference image $y$.

    \begin{equation}
        Q2^n(x,y) =
        \frac{\sigma_{x,y}}{\sigma_x\sigma_y}
        \cdot
        \frac{2\bar x \bar y}{(\bar x)^2+(\bar y)^2}
        \cdot
        \frac{2\sigma_x\sigma_y}{\sigma_x^2+\sigma_y^2}
    \end{equation}
    Here $\sigma_{x,y}$ is the covariance between $x$ and $y$, and $\sigma_x$ and $\bar x$ are the standard deviation and mean of $x$, respectively. The $Q2^n$ metric represents a good candidate to give an overall evaluation of both radiometric and spectral distortions in the pansharpened images.
    
\end{itemize}

\subsection{Full Resolution Metrics}

For the FR assessment, we decided to use the \textit{Quality with No Reference} index ($QNR$), as done by \cite{vivone2022panchromatic}. This index is obtained as the product of the spectral distortion index $D_{\lambda}$  and the spatial distortion index $D_s$.

The spectral distortion index $D_{\lambda}$ is computed as proposed in the Filtered-based QNR (FQNR) quality index \citep{arienzo2022full}. In this definition, each fused HS band is spatially degraded using its specific Modulation Transfer Function (MTF) matched filter\footnote{The filter is defined for ensuring the consistency property of the Wald's protocol \citep{wald1997fusion}. As done by \cite{vivone2022panchromatic}, we used the assumption that the HS sensor’s MTFs follow a Gaussian shape with a standard deviation set all equal to 0.3.}, then the $Q2^n$ index between the set of spatially degraded HS images and the  set of original HS data is computed, and eventually the unit complementary value is taken in order to obtain a distortion measure:

\begin{equation}
    D_{\lambda} = 1 - Q2^n( \hat H_{L\downarrow}, H )
\end{equation}

where, $\hat H_{L\downarrow}$ is the pansharpened image which has been spatially degraded using the MTF filter and decimated to input spatial dimension and $H$ are the input hyperspectral bands. As done by \cite{vivone2022panchromatic}, we adopted the $Q$ ($UQI$) index instead of the $Q2^n$ index for computational reasons due to the high number of HS bands. As stated by \cite{vivone2022panchromatic}, comparable performance can be obtained with this modification, while drastically improving the computation time.

Spatial consistency $D_s$ is computed as described by \cite{alparone2018spatial}. Adopting a linear regression framework, the PAN image is modeled as a linear combination of the fused HS bands. To measure the extent of the spatial matching between the fused HS bands and the PAN image, the coefficient of determination is exploited \citep{alparone2018spatial}.
\begin{equation}
    D_s = 1 - R^2
\end{equation}

\noindent Finally, the $QNR$ index is calculated as:
\begin{equation}
    QNR = (1 - D_\lambda)^\alpha \cdot (1 - D_s)^\beta
\end{equation}
Here the two exponents $\alpha$ and $\beta$ determine the non-linearity of response in the interval $[0,1]$
. The value of these two parameters has been set to 1, based on previous work choices \citep{vivone2022panchromatic}.

\subsection{Methods}

We compared six deep learning and three traditional {machine-learning-free} approaches,
{The selection of the methods has been done taking in consideration two factors: how recent is the approach and the availability of the source code.}
For what concerns the machine-learning-free approaches, we have chosen Principal Component Analysis (PCA)~\citep{chavez1991comparison}, Gram-Schmidt Adaptive (GSA)~\citep{4305344} and HySure~\citep{simoes2014convex}. For all these methods, we used the implementation available in the \textsc{Mini Toolbox PRISMA}
\footnote{\url{https://openremotesensing.net/wp-content/uploads/2022/11/Mini-Toolbox-PRISMA.zip}}.
Regarding the deep learning methods, we selected PNN~\citep{masi2016pansharpening}, PanNet~\citep{yang2017pannet}, MSDCNN~\citep{yuan2018multiscale}, TFNet~\citep{liu2020remote}, SRPPNN~\citep{cai2020super} and DIPNet~\citep{xie2021detail}.

Since we are interested in the evaluation of the $6\times$ upscaling pansharpening task, we modified the methods originally designed for scale factors power of 2 (e.g. $2\times$, $4\times$, $8\times$ etc...). These methods are:

\begin{itemize}

    \item \textbf{SRPPNN~\citep{cai2020super}:} the architecture proposed by \cite{cai2020super} is characterized by multiple progressive upsampling steps, that correspond to a first $2\times$ and a secondary latter $4\times$ upscaling operations. We changed those two upsampling operations by modifying the scale factors to $3\times$ and $6\times$ respectively. The rest of the original architecture has not been changed.
    
    \item \textbf{DIPNet~\citep{xie2021detail}:} this model is made of 3 main components. 
    The first two are feature extraction branches, respectively for the low-frequency and high-frequency details of the panchromatic image; here, we changed the stride value of the second convolutional layer used to reduce the features' spatial resolution, from 2 to 3, in order to bring the extracted features at the same dimension of the input bands to perform feature concatenation.
    The third component is the main branch, which uses the features extracted from the previous components along with the input images to perform the actual pansharpening operation. The main branch can be also divided into two other components: a first upsampling part and an encoder-decoder structure for signal post-processing. 
    We changed the scaling factor of the upscaling module from 2 to 3, and in the encoder-decoder part we changed the stride values of the central convolutional and deconvolutional layers from 2 to 3. 
\end{itemize}

Each method has been retrained on the proposed PRISMA dataset (RR version) by using a workstation equipped with a Titan V GPU and Ubuntu 22.04 Operating System. The environment for the training has been written in PyTorch \texttt{v1.10.0}. For all the methods the training process lasted $1000$ epochs, with a  learning rate of $1e^{-4}$ and the Adam optimizer. The loss functions used are the ones adopted in the original papers of each method.

\section{Results}

\subsection{Quantitative Comparison}

Table \ref{tab:comparisonRR} and Table \ref{tab:comparisonFR} report the numerical results of the different selected approaches with the RR and FR protocols respectively. 

\begin{table}
    \centering
    \caption{Results of the methods for the Reduced Resolution (RR) protocol. The dimensions (millions of parameters) of each model are reported alongside the results.}
    \label{tab:comparisonRR}
    \adjustbox{width=\textwidth}{
    \begin{tabular}{lcrrrr}
    \hline
    Method & \multicolumn{1}{l}{\# of parameters (M)} & \multicolumn{1}{l}{ERGAS~$\downarrow$} & \multicolumn{1}{l}{SAM~$\downarrow$} & \multicolumn{1}{l}{SCC~$\uparrow$}  & \multicolumn{1}{l}{$Q2^n~\uparrow$} \\\hline

    PCA~\citep{chavez1991comparison} & - & 8.9545 & 4.8613 & 0.6414  & 0.6071 \\
    GSA~\citep{4305344} & - & 7.9682 & 4.3499 & 0.6642 & 0.6686 \\
    HySure~\citep{simoes2014convex} & - & 8.3699 & 4.8709 & 0.5832  & 0.5610 \\\hline
    PNN~\citep{masi2016pansharpening}    & 0.08 & 12.8840                   & 3.8465                  & 0.8237                  & 0.6702                  \\
    PanNet~\citep{yang2017pannet} & 0.19 & 6.7062                    & 2.7951                  & 0.8705                  & 0.7659                  \\
    MSDCNN~\citep{yuan2018multiscale} & 0.19 & 9.9105                    & 3.0733                  & 0.8727                  & 0.7537                  \\
    TFNet~\citep{liu2020remote}  & 2.36 & \underline{6.4090}                    & 2.4644                  & \underline{0.8875}      & \underline{0.7897}      \\
    SRPPNN~\citep{cai2020super} & 1.83 & 6.4702        & \underline{2.3823}      & \textbf{0.8890}         & 0.7708                  \\
    DIPNet\citep{xie2021detail} & 2.95 & \textbf{5.1830}           & \textbf{2.3715}         & 0.8721                  & \textbf{0.7929}         \\\hline
    \end{tabular}
    }
\end{table}

\begin{figure}
    \centering

    \includegraphics[width=\textwidth]{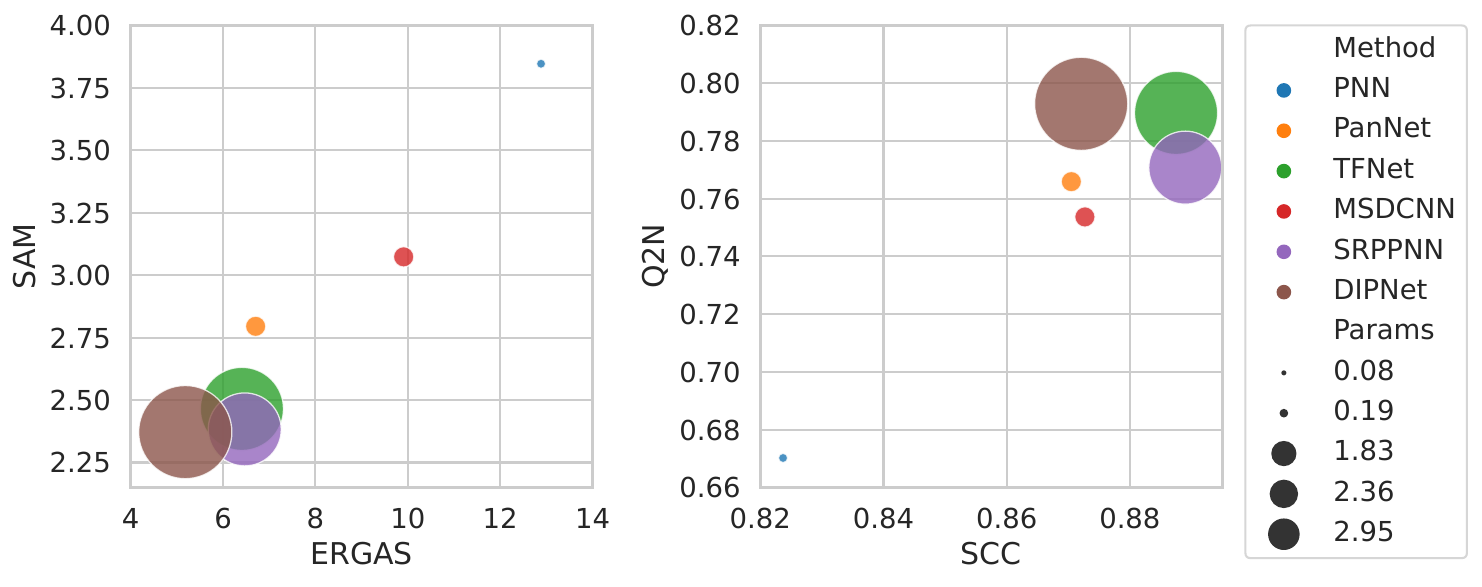}

    \caption{Graph comparison of the results of the analyzed methods with the RR protocol. The larger is the size of the circle the higher is the number of parameters (measured in millions).}
    \label{fig:graph_comparison}
\end{figure}

The two best methods for RR pansharpening protocol are DIPNet and TFNet. In-depth analysis reveals that DIPNet outperforms all other approaches across various metrics, except for the SCC index, where it ranks fourth. As the second-best algorithm, TFNet demonstrates commendable results that are comparable to those achieved by SRPPNN. Notably, machine-learning-free approaches generally exhibit lower performance compared to the majority of neural-network-based methods.

Figure \ref{fig:graph_comparison} reports a graphical comparison between network-based approaches (in the RR protocol). The comparison evaluates the performance in terms of ERGAS versus SAM (Figure \ref{fig:graph_comparison}a) and SCC versus $Q2^n$ (Figure \ref{fig:graph_comparison}b), along with the number of parameters associated with the neural models. The size of the circles in the figure corresponds to the number of parameters, measured in millions. Larger circles indicate a higher number of parameters. Ideally, the optimal approach would be represented by a small circle positioned in the bottom-left part of Figure \ref{fig:graph_comparison}a and the top-right part of Figure \ref{fig:graph_comparison}b. In practice, the best neural methods are DIPNet, SRPPNN, and TFNet which have a number of parameters that is about 30 times the number of parameters of less-performing approaches, such as PanNet and MSDCNN.

The results obtained in the FR protocol are presented in Table~\ref{tab:comparisonFR}, revealing a significant shift in the behavior of the models. Notably, TFNet emerges as the top-performing model in terms of $QNR$ index. Surprisingly, DIPNet, which was the winning method in the RR protocol, demonstrates considerably poorer results compared to other approaches. Even the simpler and smaller PanNet outperforms DIPNet, securing the second position in the comparison. 

Analyzing the spatial distortion aspect ($D_s$), the top-performing models are TFNet and MSDCNN, while DIPNet exhibits the weakest performance among the deep learning models. {It is worth mentioning that HySure is the best method in terms of $D_s$; however, additional insights regarding its performance can be found in Section~\ref{sec:qualitative_analysis}, showing various issues in the spatial reconstruction of this technique.}

On the other hand, from a spectral distortion perspective ($D^k_\lambda$), PanNet emerges as the best approach, followed by TFNet and DIPNet. A comprehensive qualitative comparison of these two aspects of the reconstruction is presented in the subsequent section. Notably, PanNet's achievement of the second-best position in the FR leaderboard is particularly noteworthy, given its comparatively smaller size compared to TFNet and other more recent approaches.

\begin{table}[t]
    \centering
    \caption{Results of the methods for the Full Resolution (FR) protocol. The dimensions (millions of parameters) of each model are reported alongside the results.
    }
    \label{tab:comparisonFR}
    \adjustbox{width=\linewidth}{
    \begin{tabular}{lcrrr}
    \hline
    Method & \multicolumn{1}{l}{\# of parameters (M)}& \multicolumn{1}{l}{$D_{\lambda} ~\downarrow$} & \multicolumn{1}{l}{$D_s~ \downarrow$} & \multicolumn{1}{l}{$QNR~ \uparrow$} \\\hline
    PCA~\citep{chavez1991comparison} & - & 0.9411 & 1.5277 & 0.0558 \\
    GSA~\citep{4305344} & - &  0.3820 & 0.0016 & 0.6170 \\
    HySure~\citep{simoes2014convex} & - & 0.4151 & 0.0009 & 0.5843 \\\hline
    PNN~\citep{masi2016pansharpening}    & 0.08 & 0.3801 & 0.0101 & 0.6136 \\
    PanNet~\citep{yang2017pannet}  & 0.19 & \textbf{0.3507} & 0.0203 & \underline{0.6360} \\
    MSDCNN~\citep{yuan2018multiscale} & 0.19 & 0.3915 & \underline{0.0068} & 0.6044 \\
    TFNet~\citep{liu2020remote} & 2.36 & \underline{0.3552} & \textbf{0.0066} & \textbf{0.6405} \\
    SRPPNN~\citep{cai2020super}  & 1.83 & 0.3948 & 0.0139 & 0.5965 \\
    DIPNet~\citep{xie2021detail}  & 2.95 & 0.3681 & 0.0348 & 0.6098 \\\hline
    \end{tabular}
    }
\end{table}

In conclusion, TFNet emerges as the most successful approach when evaluating both the RR and FR protocols. Notably, TFNet exhibits a commendable ability to strike a balance between preserving spectral and spatial information throughout the pansharpening process, particularly evident in the FR test case. When compared to SRPPNN and DIPNet, TFNet demonstrates superior generalization capabilities when transitioning from the training resolution of $180~m/px$ to the native resolution of $30~m/px$ of the PRISMA satellite hyperspectral images.

\subsection{Qualitative Comparison}
\label{sec:qualitative_analysis}

In this section, we present a qualitative comparison of the results on a selection of test images, analyzing the results in terms of the preservation of spatial and spectral distortions after the pansharpening process.

\begin{figure}
    \centering
    \begin{subfigure}[b]{0.19\textwidth}
     \centering
     \includegraphics[width=\textwidth]{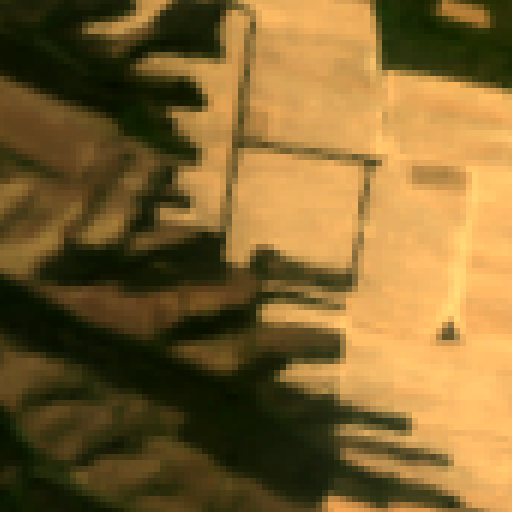}     
    \end{subfigure}
    \begin{subfigure}[b]{0.19\textwidth}
     \centering
     \includegraphics[width=\textwidth]{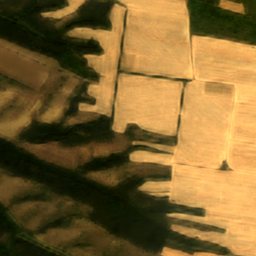}
    \end{subfigure}
    \begin{subfigure}[b]{0.19\textwidth}
     \centering
     \includegraphics[width=\textwidth]{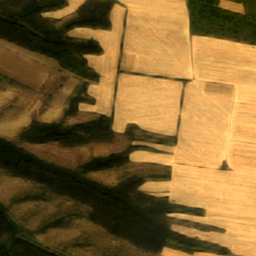}
    \end{subfigure}
    \begin{subfigure}[b]{0.19\textwidth}
     \centering
     \includegraphics[width=\textwidth]{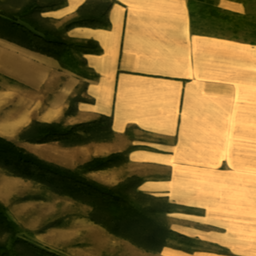}
    \end{subfigure}
    \begin{subfigure}[b]{0.19\textwidth}
     \centering
     \includegraphics[width=\textwidth]{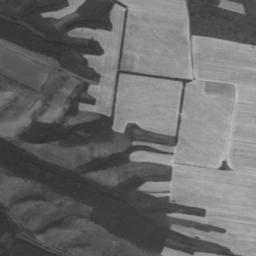}
    \end{subfigure}
    
    \begin{subfigure}[b]{0.19\textwidth}
     \centering
     \includegraphics[width=\textwidth]{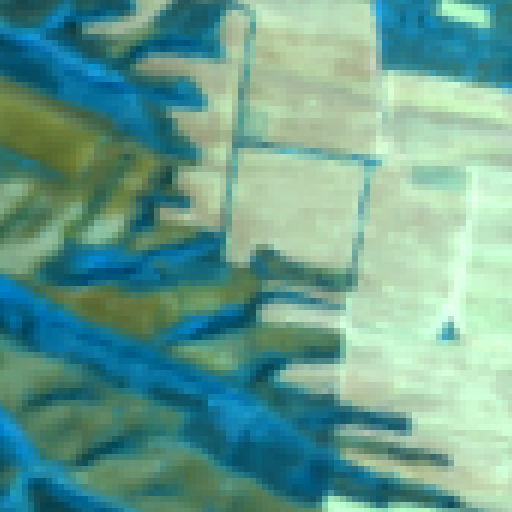}
     \caption{Input \tiny{(30m/px)}}    
    \end{subfigure}
    \begin{subfigure}[b]{0.19\textwidth}
     \centering
     \includegraphics[width=\textwidth]{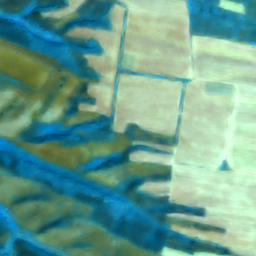}
     \caption{DIPNet \tiny{(5m/px)}}
    \end{subfigure}
    \begin{subfigure}[b]{0.19\textwidth}
     \centering
     \includegraphics[width=\textwidth]{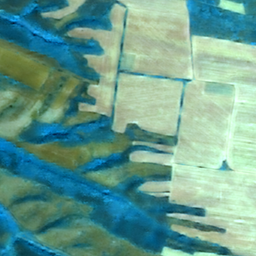}
     \caption{PanNet \tiny{(5m/px)}}
    \end{subfigure}
    \begin{subfigure}[b]{0.19\textwidth}
     \centering
     \includegraphics[width=\textwidth]{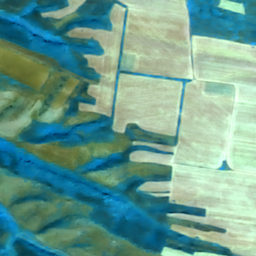}
     \caption{TFNet \tiny{(5m/px)}}
    \end{subfigure}
    \begin{subfigure}[b]{0.19\textwidth}
     \caption{PAN \tiny{(5m/px)}}
    \end{subfigure}
    
    \caption{Pansharpening results on a $512 \times 512$ tile of a test set image. For visualization purposes, images have been linearly stretched between the 1 and 99 percentile of the image histogram. In the first row, images are visualized in true color ($641~nm$, $563~nm$, $478~nm$), and in the second row, images are in false color ($1586~nm$, $1229~nm$, $770~nm$).}
    \label{fig:visualcomp1}
\end{figure}

\begin{figure}
   \begin{subfigure}[b]{0.19\textwidth}
    \centering
    \includegraphics[width=\textwidth]{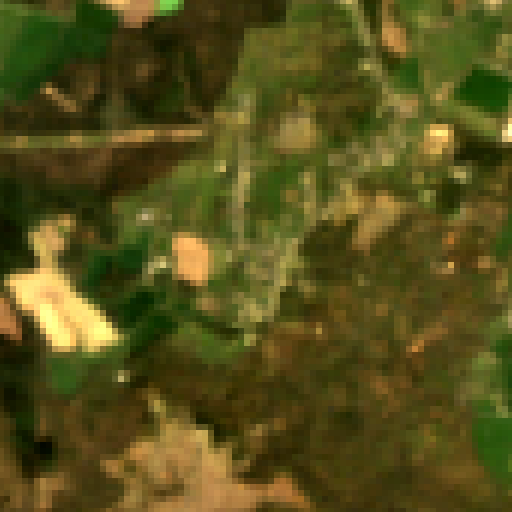}
   \end{subfigure}
   \begin{subfigure}[b]{0.19\textwidth}
    \centering
    \includegraphics[width=\textwidth]{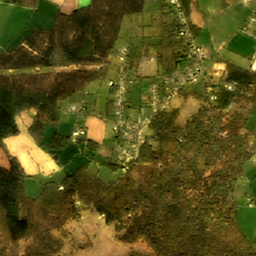}
   \end{subfigure}
    \begin{subfigure}[b]{0.19\textwidth}
     \centering
     \includegraphics[width=\textwidth]{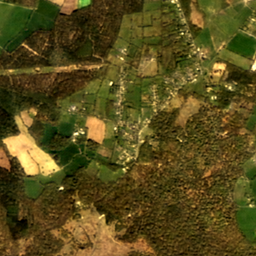}

    \end{subfigure}
    \begin{subfigure}[b]{0.19\textwidth}
     \centering
     \includegraphics[width=\textwidth]{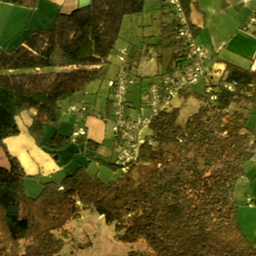}
    \end{subfigure}
   \begin{subfigure}[b]{0.19\textwidth}
    \centering
    \includegraphics[width=\textwidth]{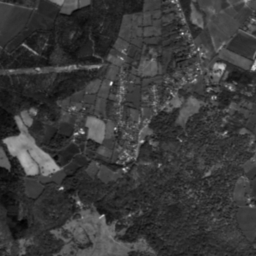}
   \end{subfigure}

   \begin{subfigure}[b]{0.19\textwidth}
    \centering
    \includegraphics[width=\textwidth]{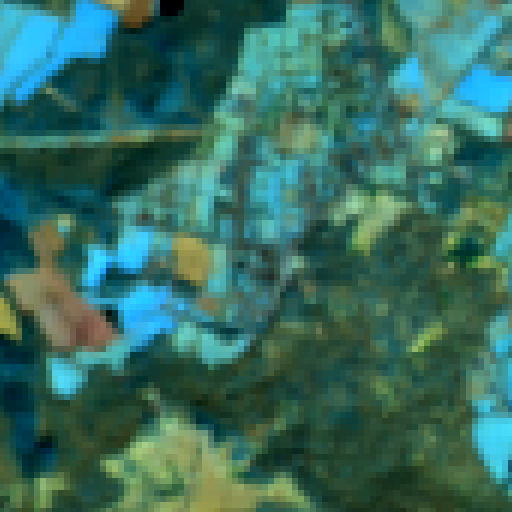}
    \caption{Input \tiny{(30m/px)}}
   \end{subfigure}
   \begin{subfigure}[b]{0.19\textwidth}
    \centering
    \includegraphics[width=\textwidth]{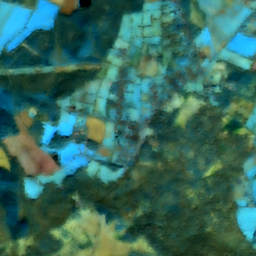}
    \caption{DIPNet \tiny{(5m/px)}}
   \end{subfigure}
    \begin{subfigure}[b]{0.19\textwidth}
     \centering
     \includegraphics[width=\textwidth]{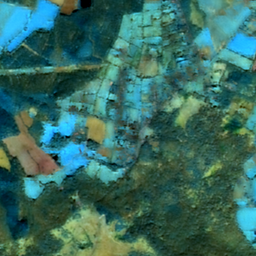}
     \caption{PanNet \tiny{(5m/px)}}
    \end{subfigure}
    \begin{subfigure}[b]{0.19\textwidth}
     \centering
     \includegraphics[width=\textwidth]{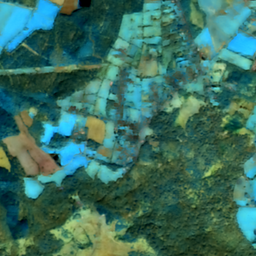}
     \caption{TFNet \tiny{(5m/px)}}
    \end{subfigure}
   \begin{subfigure}[b]{0.19\textwidth}
    \centering
    \caption{PAN \tiny{(5m/px)}}
   \end{subfigure}
    \caption{Pansharpening results on a $512 \times 512$ tile of a test set image. For visualization purposes, images have been linearly stretched between the 1 and 99 percentile of the image histogram. In the first row, images are visualized in true color ($641~nm$, $563~nm$, $478~nm$), and in the second row, images are in false color ($1586~nm$, $1229~nm$, $770~nm$).}
    \label{fig:visualcomp2}
\end{figure}

\begin{figure}
    \centering
    \begin{subfigure}[b]{0.19\textwidth}
     \centering
     \includegraphics[width=\textwidth]{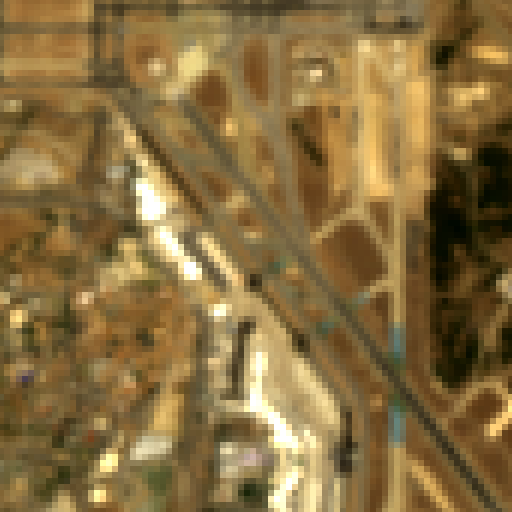}
     
    \end{subfigure}
    \begin{subfigure}[b]{0.19\textwidth}
     \centering
     \includegraphics[width=\textwidth]{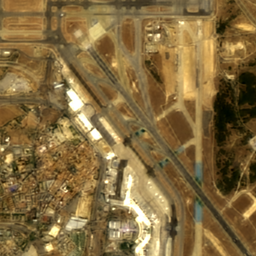}
     
    \end{subfigure}
    \begin{subfigure}[b]{0.19\textwidth}
     \centering
     \includegraphics[width=\textwidth]{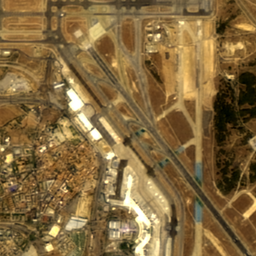}
     
    \end{subfigure}
    \begin{subfigure}[b]{0.19\textwidth}
     \centering
     \includegraphics[width=\textwidth]{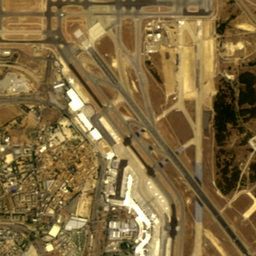}  
    \end{subfigure}
    \begin{subfigure}[b]{0.19\textwidth}
     \centering
     \includegraphics[width=\textwidth]{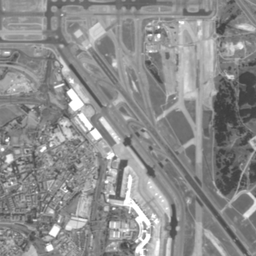}
    \end{subfigure}
    
    \begin{subfigure}[b]{0.19\textwidth}
     \centering
     \includegraphics[width=\textwidth]{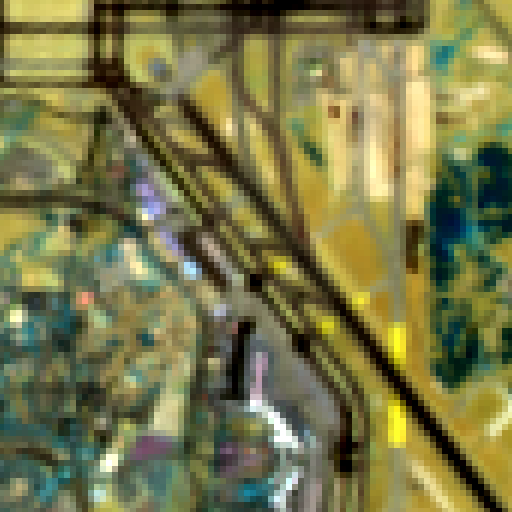}
     \caption{Input \tiny{(30m/px)}}
    \end{subfigure}
    \begin{subfigure}[b]{0.19\textwidth}
     \centering
     \includegraphics[width=\textwidth]{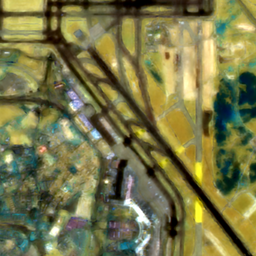}
     \caption{DIPNet \tiny{(5m/px)}}
    \end{subfigure}
    \begin{subfigure}[b]{0.19\textwidth}
     \centering
     \includegraphics[width=\textwidth]{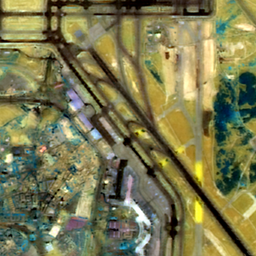}
     \caption{PanNet \tiny{(5m/px)}}
    \end{subfigure}
    \begin{subfigure}[b]{0.19\textwidth}
     \centering
     \includegraphics[width=\textwidth]{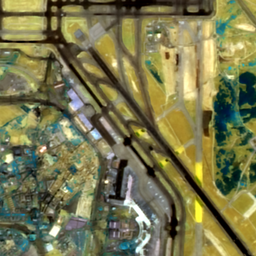}
     \caption{TFNet \tiny{(5m/px)}}
    \end{subfigure}
    \begin{subfigure}[b]{0.19\textwidth}
     \centering
     \caption{PAN \tiny{(5m/px)}}
    \end{subfigure}

    \caption{Pansharpening results on a $512 \times 512$ tile of a test set image. For visualization purposes, images have been linearly stretched between the 1 and 99 percentile of the image histogram. In the first row, images are visualized in true color ($641~nm$, $563~nm$, $478~nm$), and in the second row, images are in false color ($1586~nm$, $1229~nm$, $770~nm$).}
    \label{fig:visualcomp3}
\end{figure}

Figures \ref{fig:visualcomp1}, \ref{fig:visualcomp2}, and \ref{fig:visualcomp3} show the results of the best models on three images of the FR protocol. Here are shown center crops of dimensions  $512\times512$ of the pansharpened images ($5~m$ per pixel) alongside the same crop of the original input image ($30~m$ per pixel). For visualization purposes, images have been linearly stretched between the 1 and 99 percentile of the image histogram. In the first row, images are visualized in true color (641 nm, 563 nm, 478 nm), and in the second row, images are in false color (1586 nm, 1229 nm, 770 nm). For what concerns the spatial information, as can be seen here and as already highlighted by the quantitative comparison, TFNet presents the overall best-looking structures and details. Among the considered methods and especially in comparison with DIPNet, TFNet reconstructs much cleaner images, with good edges and a lot more details. PanNet still achieves good results compared to DIPNet but with still few artifacts and aberrations of different kinds. In Figures \ref{fig:visualcomp2} and \ref{fig:visualcomp3}, it is easy to notice the presence of such artifacts, due to the high amount of details in the scenes. Overall, DIPNet presents the most blurry results, with a very poor amount of details and the presence of artifacts, particularly noticeable in the false color composite version of the reported scenes.

\begin{figure}
    \centering

    \begin{subfigure}[b]{0.19\textwidth}
     \centering
     \includegraphics[width=\textwidth]{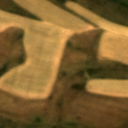}
    \end{subfigure}
    \begin{subfigure}[b]{0.19\textwidth}
     \centering
     \includegraphics[width=\textwidth]{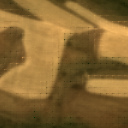}
    \end{subfigure}
    \begin{subfigure}[b]{0.19\textwidth}
     \centering
     \includegraphics[width=\textwidth]{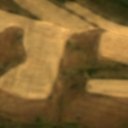}
    \end{subfigure}
    \begin{subfigure}[b]{0.19\textwidth}
     \centering
     \includegraphics[width=\textwidth]{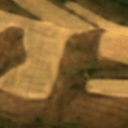}
    \end{subfigure}
    \begin{subfigure}[b]{0.19\textwidth}
     \centering
     \includegraphics[width=\textwidth]{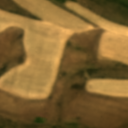}
     
    \end{subfigure}

    \begin{subfigure}[b]{0.19\textwidth}
     \centering
     \includegraphics[width=\textwidth]{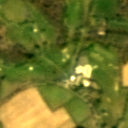}
    \end{subfigure}
    \begin{subfigure}[b]{0.19\textwidth}
     \centering
     \includegraphics[width=\textwidth]{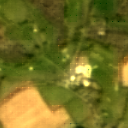}
    \end{subfigure}
    \begin{subfigure}[b]{0.19\textwidth}
     \centering
     \includegraphics[width=\textwidth]{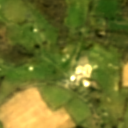}
    \end{subfigure}
    \begin{subfigure}[b]{0.19\textwidth}
     \centering
     \includegraphics[width=\textwidth]{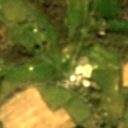}
    \end{subfigure}
    \begin{subfigure}[b]{0.19\textwidth}
     \centering
     \includegraphics[width=\textwidth]{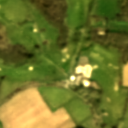}
    \end{subfigure}
    
    \begin{subfigure}[b]{0.19\textwidth}
     \centering
     \includegraphics[width=\textwidth]{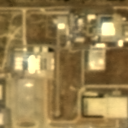}     
     \caption{GSA}
    \end{subfigure}
    \begin{subfigure}[b]{0.19\textwidth}
     \centering
     \includegraphics[width=\textwidth]{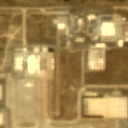}
     \caption{HySure}
    \end{subfigure}
    \begin{subfigure}[b]{0.19\textwidth}
     \centering
     \includegraphics[width=\textwidth]{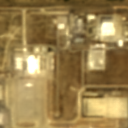}
     \caption{DIPNet}
    \end{subfigure}
    \begin{subfigure}[b]{0.19\textwidth}
     \centering
     \includegraphics[width=\textwidth]{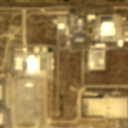}
     \caption{PanNet}
    \end{subfigure}
    \begin{subfigure}[b]{0.19\textwidth}
     \centering
     \includegraphics[width=\textwidth]{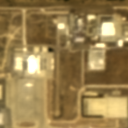}
     \caption{TFNet}
    \end{subfigure}

    \caption{Zoom of areas from the test images. Crops of dimension $128\times128$, at resolution $5~m/px$, in true color ($641~nm$, $563~nm$, $478~nm$). As can be seen, repeated artifacts along the edges can be observed for the HySure method.}
    \label{fig:artifacts}
\end{figure}

Figure~\ref{fig:artifacts} shows zoomed crops at dimension $128\times128$ of areas of the same images processed by the best neural-based and the two best machine-learning-free methods. As can be seen, even if HySure numerically represents the best approach from the spatial point of view (see Table~\ref{tab:comparisonFR}, $D_s$ index), a pattern of artifacts occur over all the images processed by the HySure algorithm.
This last comparison shows a potential problem in the adoption of $QNR$ index as a metric for the no-reference analysis, when this type of artifacts occur in the pansharpened images.

\begin{figure}[t]
    \centering

    \begin{subfigure}[]{0.49\linewidth}
        \includegraphics[width=\textwidth]{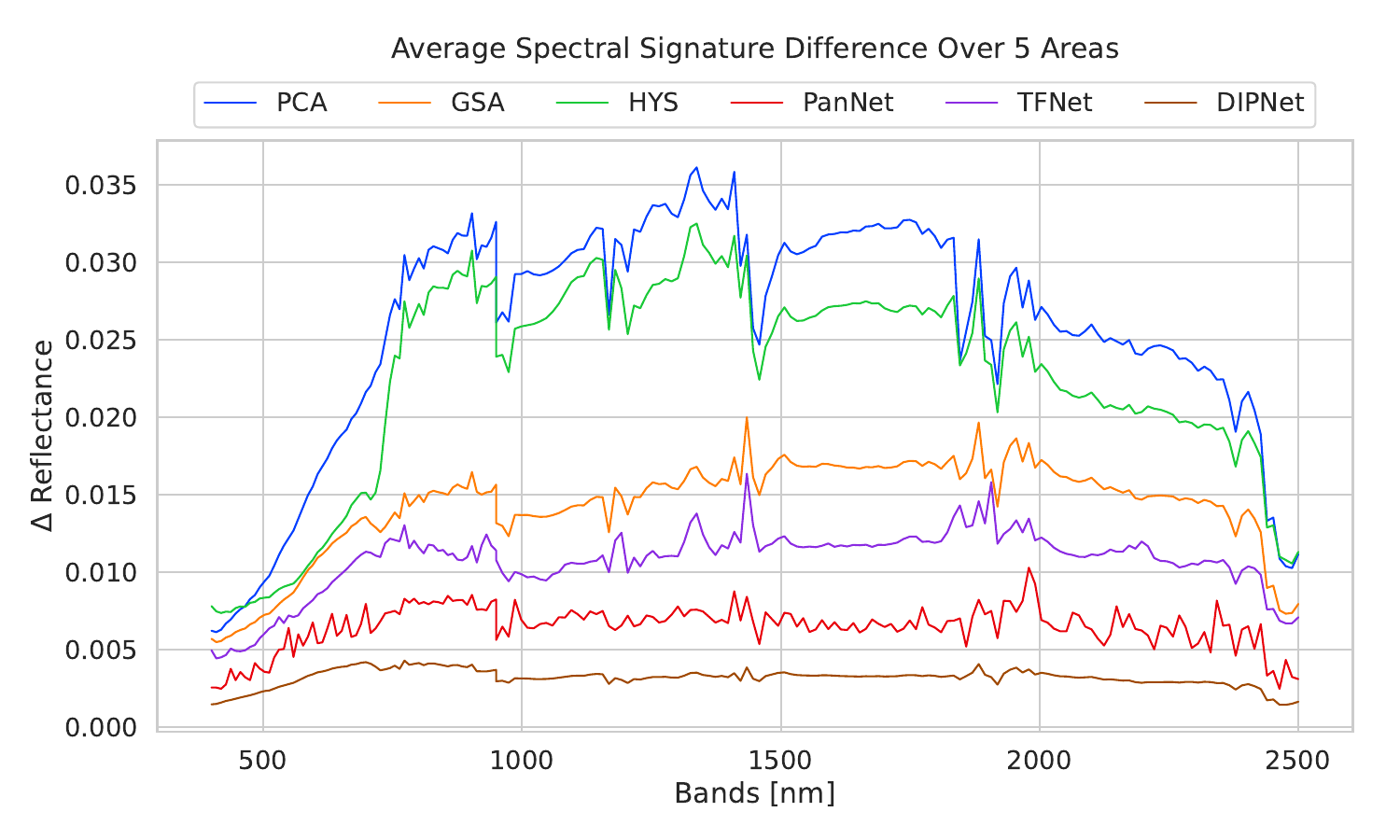}
    \end{subfigure}
    \begin{subfigure}[]{0.49\linewidth}
        \includegraphics[width=\textwidth]{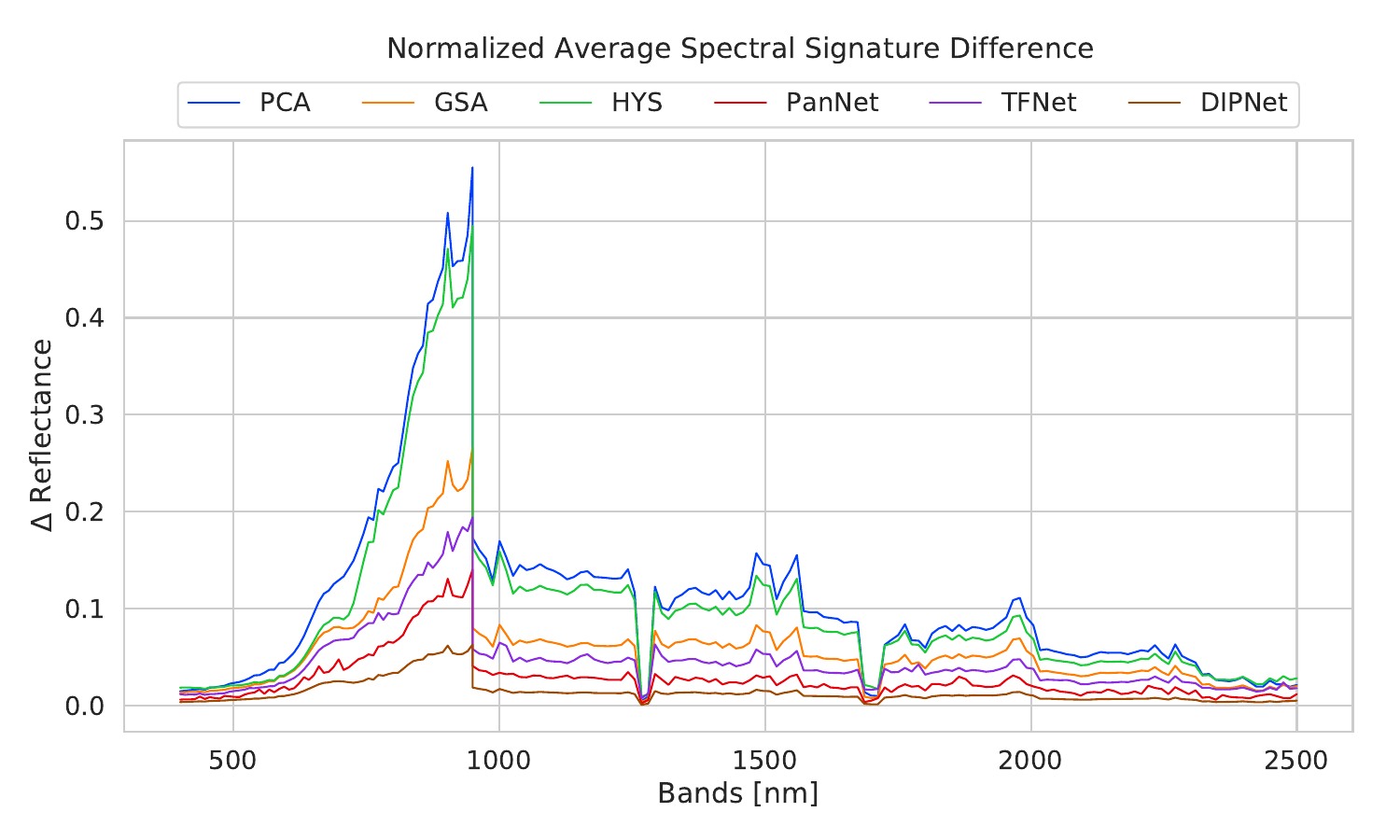}    
    \end{subfigure}

    \adjustbox{width=\textwidth}{
    \begin{subfigure}[]{0.2\linewidth}
        \includegraphics[width=\linewidth]{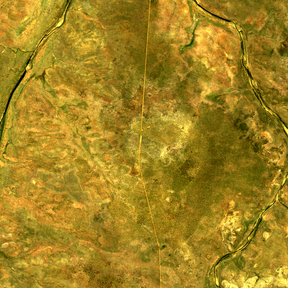}    
    \end{subfigure}
    \begin{subfigure}[]{0.2\linewidth}
        \includegraphics[width=\linewidth]{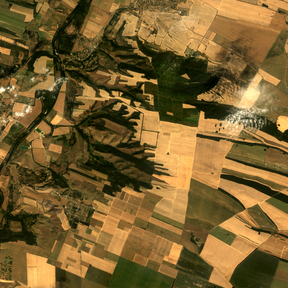}    
    \end{subfigure}
    \begin{subfigure}[]{0.2\linewidth}
        \includegraphics[width=\linewidth]{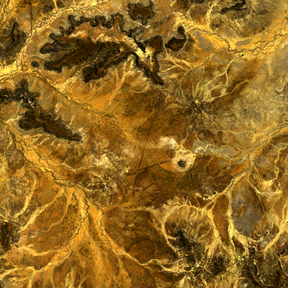}
    \end{subfigure}
    \begin{subfigure}[]{0.2\linewidth}
        \includegraphics[width=\linewidth]{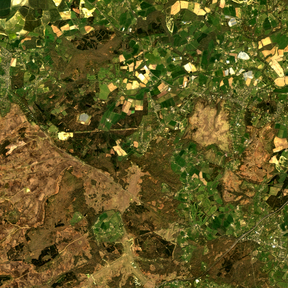}    
    \end{subfigure}
    \begin{subfigure}[]{0.2\linewidth}
        \includegraphics[width=\linewidth]{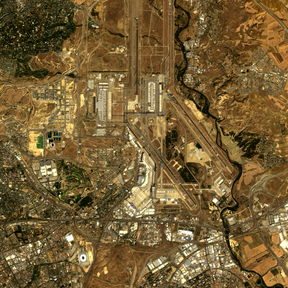}    
    \end{subfigure}
    }
    
    \caption{Difference between spectral signatures of the fused images with respect to the input image. The difference is evaluated as the average of the differences for each pixel of the five images reported in the row below the graph. The graph on the left shows the average spectral difference, while the graph on the right shows the difference normalized for each band.}
    \label{fig:MAE}
\end{figure}

\begin{figure}
    \centering
    \adjustbox{width=\textwidth}{
    \includegraphics{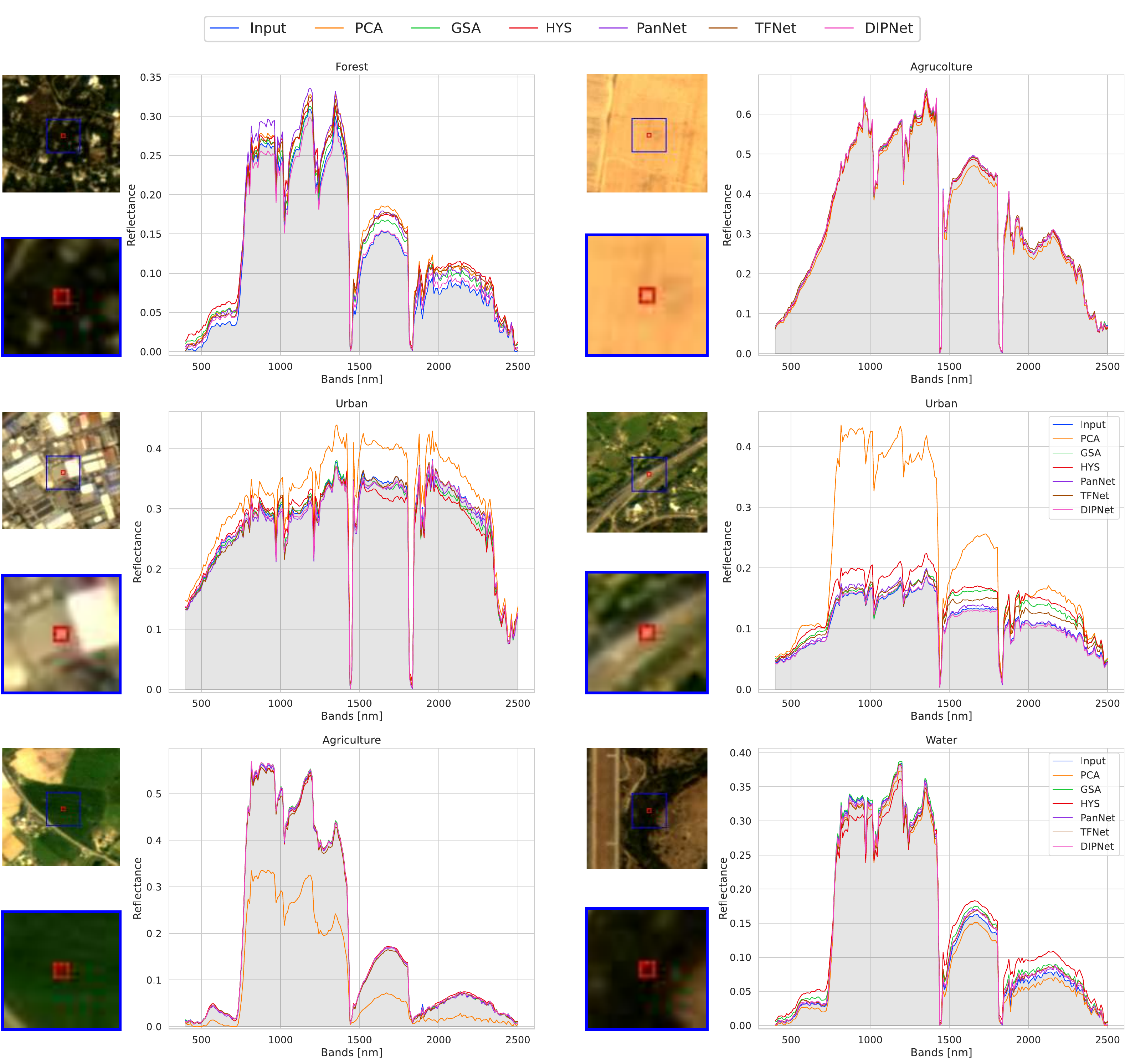}
    }
    \caption{Spectral signatures obtained in six different areas, labeled as \textit{Forest}, \textit{Urban}, \textit{Agriculture}, and \textit{Water} areas. For each area are reported the spectral signatures of the input bands and the ones obtained by each pansharpening method. The area used to extract the signatures is the one in the red box highlighted in each image. The images thumbnails are in true colors ($641~nm$, $563~nm$, $478~nm$).}
    \label{fig:signatures}
\end{figure}

Figure \ref{fig:MAE} reports the average differences between the spectral signatures of each method and the reference one, alongside the normalized version of the same difference, computed on 5 different tiles. 
These tiles have been randomly extracted from the test set. From this comparison, it is possible to notice how DIPNet's average error is much smaller with respect to the other approaches. Compared with the results from the quantitative evaluation with the FR protocol, where DIPNet reaches third place, this is the only unexpected behavior. This unexpected result could be an insight into a possible flaw in adopting the spectral component of the $QNR$ metric as it is usually done in the literature.

Figure~\ref{fig:signatures} shows the spectral signatures of both input and pansharpened images for each method. Here we considered only selected groups of pixels, specifically labeled as \textit{Forest}, \textit{Urban}, \textit{Agriculture}, and \textit{Water}, highlighted in red in the images aside the graphs. The best method is expected to show signatures closer to the input ones; we reported both the machine-learning-free and the best deep learning approaches.
From this qualitative comparison and according to the numbers in Table~\ref{tab:comparisonFR}, PCA and HySure
are the worst-performing methods in terms of spectral fidelity. For all the reported classes, these methods perform badly over the entire spectrum, and, in particular, PCA and Hysure performance is even worse than GSA.
Regarding the deep learning approaches, DIPNet seems to show the lowest difference from the input, despite the score obtained in terms of $D_{\lambda}$ (see Table~\ref{tab:comparisonFR}). TFNet and PanNet instead have a behavior more coherent with the results obtained in the quantitative evaluation with the FR protocol. PanNet seems to perform better from spectral fidelity point of view with respect to TFNet, which, however, performs better than all of the machine-learning-free approaches.
Overall, we can confirm the results in Table~\ref{tab:comparisonFR}, with DIPNet, PanNet, and TFNet as the best approaches, with performance higher in comparison to the machine-learning-free approaches.

Images at higher resolution are available at \url{https://thezino.github.io/HSbenchmarkPRISMA/}.

\section{Conclusions}

The increasing availability of hyperspectral remote sensing data presents new opportunities for studying Earth's soil. However, this type of data is typically collected at low resolution, posing challenges in their effective usability in RS tasks. Therefore, the process of image pansharpening becomes crucial in the enhancement of hyperspectral remote sensing images.

In literature, deep learning approaches have shown promising results. These methods, however, are data hungry and state-of-the-art datasets strive to support sufficient data. To overcome this limitation, we proposed a newly collected large-scale dataset using the PRISMA ASI satellite for training and assessment models.

To the best of our knowledge, this work is the first to present an analysis based on a big and various hyperspectral dataset for pansharpening in the state of the art (more than 1000 tiles, of which more than 200 are used for testing). The dataset tiles have been collected from 190 PRISMA images with 203 bands\footnote{The original number of PRISMA spectral bands is 240. The number reported above is obtained after a proposed pre-processing procedure.} covering both the VNIR and SWIR parts of the spectrum, making this investigation the optimal benchmark for hyperspectral pansharpening.

The comparison includes machine-learning-free and deep learning techniques tested using two experimental protocols for $6\times$ upscaling factor: Reduced Resolution (RR) and Full Resolution (FR). The former is used for training and testing, and the latter to test the methods on the original resolution, evaluating their ability to generalize the upsampling operation at different starting resolutions with respect to the training phase.

The RR protocol consists of a comparison between the reconstructed data and the original hyperspectral images as target references. The results show that the neural networks generally work better than the machine-learning-free methods for spatial information improvement and spectral information coherency. In particular, DIPNet and TFNet architectures outperform any other techniques evaluated.
In the FR protocol, the comparison with ground truths is not possible thus quantitative and also qualitative evaluations have been reported to have a complete understanding of the investigated methods. Based on both assessments, the architecture that achieves the best overall performance is TFNet which remains coherent with the RR results. DIPNet instead shows worse results when it comes to spatial reconstruction, not demonstrating good abilities of adaptation when the original resolution is involved and thus not being the best option for tests in real-world applications. It is also valuable to notice that machine-learning-free methods are generally worse at reconstructing the spectral information, degrading the signals. 
  
The investigation conducted in this work clearly shows that data-driven neural architectures are generally better for hyperspectral pansharpening, both in spectral and spatial reconstruction, using a dataset that allows for meaningful analyze the different approaches. On the contrary, The machine-learning-free methods are not adaptable to the new environment based on hyperspectral data and wavelengths outside the visible part of the spectrum.

In our opinion, to further improve hyperspectral pansharpening performance, future research should focus on creating new data-driven neural architectures which directly focus on hyperspectral data and the relationship between the different portions of the spectrum.

\section*{Acknowledgements}

Research developed in the context of the project PIGNOLETTO $-$ Call HUB Ricerca e Innovazione CUP (Unique Project Code) n. E41B20000050007, co-funded by POR FESR 2014-2020 (Programma Operativo Regionale, Fondo Europeo di Sviluppo Regionale – Regional Operational Programme, European Regional Development Fund).


\bibliographystyle{elsarticle-harv}
\bibliography{biblio}

\end{document}